# An integrated evanescent-field biosensor in silicon

Mohammed A. Al-Qadasi[*,1,#], Samantha M. Grist[*,1-4,#], Matthew Mitchell[4,5], Karyn Newton[1,3], Stephen Kioussis[1,3], Sheri J. Chowdhury[1,5], Avineet Randhawa[1-3], Yifei Liu[1,3], Piramon Tisapramotkul[1,3], Karen C. Cheung[1-3], Lukas Chrostowski[1,4,5], Sudip Shekhar[1,4,#]

[*] Contributed equally
[1] Department of Electrical and Computer Engineering, The University of British Columbia
[2] School of Biomedical Engineering, The University of British Columbia
[3] Centre for Blood Research, The University of British Columbia
[4] Dream Photonics, Inc.
[5] Stewart Blusson Quantum Matter Institute, The University of British Columbia
[#] Corresponding authors: alqadasi@ece.ubc.ca, sgrist@ece.ubc.ca, sudip@ece.ubc.ca

## Abstract

Decentralized diagnostic testing that is accurate, portable, quantitative, and capable of making multiple simultaneous measurements of different biomarkers at the point-of-need remains an important unmet need in the post-pandemic world. Resonator-based biosensors using silicon photonic integrated circuits are a promising technology to meet this need, as they can leverage (1) semiconductor manufacturing economies of scale, (2) exquisite optical sensitivity, and (3) the ability to integrate tens to hundreds of sensors on a millimeter-scale photonic chip. However, their application to decentralized testing has historically been limited by the expensive, bulky tunable lasers and alignment optics required for their readout. In this work, we introduce a segmented sensor architecture that addresses this important challenge by facilitating resonance-tracking readout using a fixed-wavelength laser. The architecture incorporates an in-resonator phase shifter modulated by CMOS drivers to periodically sweep and acquire the resonance peak shifts as well as a distinct high-sensitivity sensing region, maintaining high performance at a fraction of the cost and size. We show, for the first time, that fixed-wavelength sensor readout can offer similar performance to traditional tunable laser readout, demonstrating a system limit of detection of $6.1 \pm 1.9 \times 10^{-5}$ RIU as well as immunoassay-based detection of the SARS-CoV-2 spike protein. We anticipate that this sensor architecture will open the door to a new data-rich class of portable, accurate, multiplexed diagnostics for decentralized testing.

## Main

Biomarker-based testing, most of which is performed in centralized testing laboratories, provides critical evidence for physicians' diagnostic decisions. Although this type of laboratory testing is the gold standard for accuracy and facilitates simultaneous testing for several biomarkers, there remain significant challenges with turnaround time, access to testing, and sample transport[1] – these challenges create an urgent need for portable, low-cost, accurate, and widely available diagnostics. This need was underscored by the COVID-19 pandemic[2,3], but there exists a suite of needs for decentralized testing, including time-sensitive treatment decisions in cardiac diagnostics[4] and traumatic brain injury[5], diagnostics in remote and rural communities[6], and monitoring health conditions at home[1]. Current rapid, portable diagnostics like lateral flow assays, however, do not provide the kind of robust, quantitative, and accurate results for multiplexed (multiple biomarker) detection that physicians rely on from laboratory assays[7,8].

Optical evanescent field sensors, including surface plasmon resonance (SPR) and silicon photonics (SiP)-based sensors, are one promising solution that has been proposed to address open needs for point-of-care (PoC) and decentralized biomarker detection[9–13]. By facilitating label-free detection of the intrinsic optical properties of the biomarker of interest, these technologies overcome inherent challenges of label-based detection methods (e.g., fluorescence and absorbance-based lateral flow assays and enzyme-linked immunosorbent assays (ELISAs)), including reagent selection and pairing, unintended effects of the label, and slow preparation and operation[14,15]. Technologies based upon silicon photonic (SiP) integrated circuits[16] offer particular advantages for decentralized testing because this technology leverages semiconductor manufacturing economies of scale for low-cost, high-volume production[17] and is capable of integrating multiple compact sensors and technologies on a single chip[18], including complementary sensors (e.g., temperature references) and CMOS electronics for readout[10,15]. Each silicon or silicon nitride waveguide-based sensor can have its own chemistry for multiplex detection[19], enabling many high-sensitivity assays in a single compact device[20].

SiP sensors based on optical resonator structures can measure changes in optical refractive index at the surface of the sensor with exquisite precision[21]. To use this type of sensor for detection of biomarker targets of interest, the SiP waveguides are typically functionalized with specific detection receptor molecules (that bind specifically to the biomarker target of interest) and subsequently exposed to the fluidic sample containing the biomarker analytes[22,23]. The magnitude of the resonance peak shifts after exposure to the sample corresponds to the overlap of the evanescent field with binding sites on the surface of the waveguides as well as the amount of biomarker that attaches to the waveguides[24] (which is in turn dependent on the concentration of biomarker, the affinity and kinetics of the binding, and the transport of biomarker to the surface of the sensor[25]). By leveraging these interactions between the evanescent field and biomarker, SiP biosensors enable real-time, label-free measurement of biomarkers[26–30].

Signal readout from SiP resonators is typically performed by using a tunable laser source (TLS) to interrogate the optical transmission spectrum of resonant structures by sweeping the wavelength of the coupled light[17] (Fig. 1a). However, the need for a TLS and opto-mechanical apparatus for steering and coupling the light introduces significant challenges for decentralized diagnostics due to the cost and size of the readout system (Fig. 1a). Cost-effective optical sources have been proposed as alternatives to tunable lasers, but they have several shortcomings[31,32]. In one proposed sensing-tracking resonator configuration[31], a broadband optical source is used alongside a second resonator, coupled in series to the sensing resonator, which scans for the resonance shifts of the sensing resonator. This setup incorporates multiple coupling points which lead to excess loss and degrade sensor performance. Additionally, extra components, such as optical filters, are required to mitigate undesired signals and prevent false detection[31].

In contrast to TLS, fixed-wavelength (FW) laser sources like distributed-feedback (DFB) lasers are considerably smaller (<1 mm) and less expensive (<$1 at high volume) and can be integrated into silicon photonic chips through advanced packaging techniques like photonic wire bonding[33]. Modulating the current of these lasers is an alternate approach to achieve tunability[32,33]; however, this method offers limited tuning range and introduces complex feedback control techniques for amplitude control and frequency linearization. Intensity interrogation using a fixed input wavelength has been employed as another method for overcoming the readout challenge for SiP biosensors[34]. This approach necessitates careful biasing of a resonator or interferometer at the maximum transmission slope, allowing perturbations in the waveguide medium to translate into detectable intensity fluctuations. However, sensing events that produce a large amount of signal suffer from nonlinearity and sensitivity fading[35]. Furthermore, amplitude noise from optical fluctuations directly impacts the detected signal, compromising the measurement's accuracy and limit of detection. Amplitude noise can arise from sources such as variable coupling due to mechanical vibrations, relative intensity noises of the input laser or optical amplifier, and the readout noise, including thermal and shot noise[36]. To address the sensitivity fading issue, coherent detection has been explored to measure the phase shift in the sensing arm rather than the intensity as a function of the changes in the RI[37]. This implementation incorporates a reference arm and introduces an optical hybrid which results in a larger footprint, limiting the capability for multiplexing. Additionally, it remains susceptible to amplitude noise and requires digital signal processing at the output readout to extract the phase. Phase interrogation using microring resonators has been explored as another approach of coherent detection to reduce intensity noise; however, the interrogated phase has a nonlinear relationship with the refractive index, leading to phase sensitivity fading[38].

In this paper, we address these challenges by introducing a novel sensor architecture compatible with resonance-tracking (rather than intensity) interrogation using inexpensive fixed-wavelength lasers (Fig. 1). This architecture (Fig. 1(b)) uses a segmented ring resonator-based (SRR) biosensor design, encompassing both a distinct phase-tuning region and a separated sensing region. This architecture permits resonance peak interrogation by allowing the resonator's transmission spectrum to be controllably shifted to sweep the resonance peaks past the wavelength of the readout laser – analogous to the controlled sweeping of a TLS wavelength across the resonator's transmission spectrum. In this way, the segmented sensor architecture permits resonance peak interrogation, mitigating the challenges with amplitude noise pervasive in many fixed-wavelength readout approaches. The architecture separates the tuning and sensing regions, thus effectively mitigating the effect of phase shifter-induced temperature changes that might impact the binding interactions used for specific analyte detection (Fig 1b). In addition, the design is tolerant to process variations and does not require pre-fabrication critical coupling optimization. To work towards a fully portable system, we also present a custom CMOS chip for sensor readout, providing both the thermal phase shifter sweep signals via pulse-density modulation (PDM) as well as trans-impedance amplifiers (TIAs) with adjustable gain for detection signal processing. We integrate this new sensor with a microfluidic device and validate its resonance-tracking interrogation scheme by characterizing important metrics of resonator sensor performance such as its quality factor, bulk refractive index sensitivity, noise levels, drift rate, replicability, and intrinsic and system limits of detection. Finally, we demonstrate the sensor's utility for biomarker detection using a demonstration binding assay to detect the spike protein of the SARS-CoV-2 virus that causes COVID-19. These demonstrations highlight how the modular design of our novel SRR sensor architecture yields sensing performance similar to traditional resonator sensors but at a fraction of the full-system size and cost, opening the door to decentralized, data-rich SiP diagnostics.

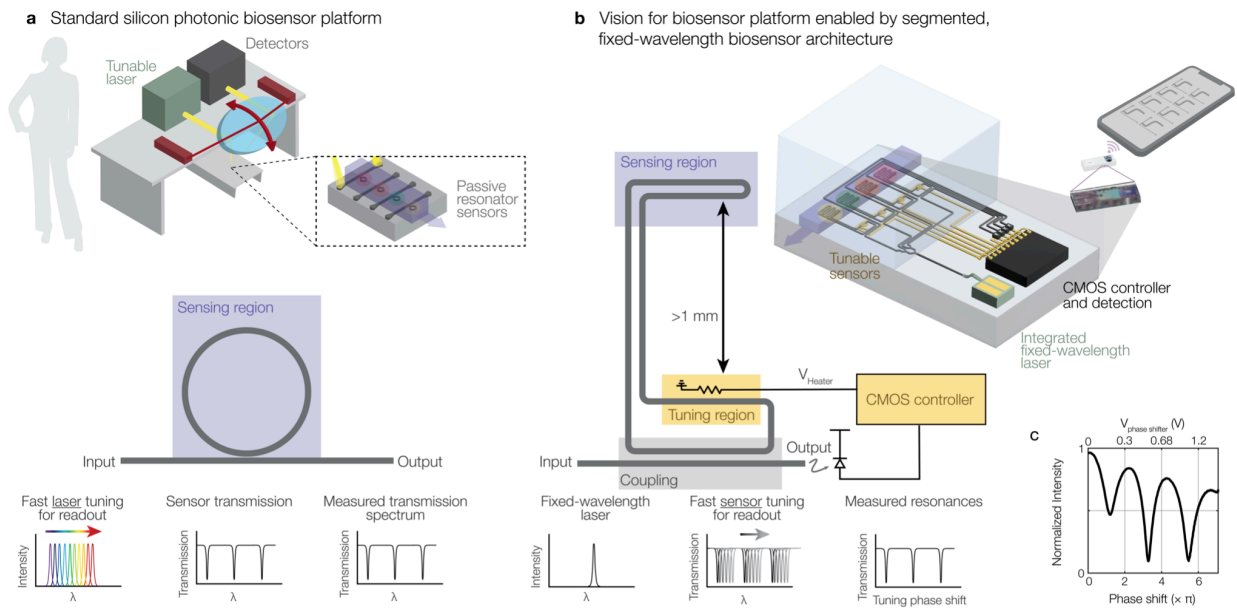

**Fig. 1 The novel fixed-wavelength sensor architecture will open the door to point-of-need silicon photonic diagnostics by permitting fully integrated readout of shifts in resonance wavelength using a fixed-wavelength laser.** In comparison to standard passive resonator sensors (a), the segmented fixed-wavelength resonator architecture (b) will enable fully integrated on-chip miniaturization of the silicon photonics readout system – a key barrier to using silicon photonic biosensors at the point of need. (a-b) depict a comparison of the full sensor readout system (top), the architecture of the on-chip silicon photonic sensors (middle), and the readout process (bottom). In the segmented architecture (b), the sensing region of the resonators is functionalized for specific detection of biomarkers of interest, and the sensor transmission function – rather than the input laser – is tuned to read out shifts in resonance wavelength. By supplying controlled resonator phase shifts using a CMOS controller, it is possible to read out resonance shifts due to analyte binding. The sensing and tuning regions are separated by >1 mm to avoid any thermal or electrical effects on the analytes or sensor functionalization. The custom CMOS readout chip facilitates both controlled sensor tuning for readout and signal amplification at the optical detector. By supplying PDM signal to the on-chip phase shifter, the sensor transmission function can be rapidly tuned across the wavelength of a fixed-wavelength laser input, permitting readout of resonances and conversion to wavelength without any tuning of the input laser. (c) Experimental resonance peaks read out using the fixed-wavelength sensor, showing transmitted optical intensity vs. phase shifter tuning voltage and resulting phase shift. Sensor architecture schematics are not drawn to scale.

## Results

**The segmented ring resonator architecture is composed of multiple purpose-built segments**

In order to introduce a sensor compatible with fixed-wavelength readout of resonance peak position, our novel SRR architecture uses a modular design with segments optimized for specific functional purposes (Fig. 2). The first segment of the SRR is the sensing region, which interfaces with the sample to facilitate specific biomolecular detection. This segment is tailored to exhibit high susceptibility to refractive index perturbations in the cladding (Fig. 2a) by leveraging sub-wavelength grating (SWG) waveguides, which are shown to enhance sensitivity due to the substantial modal overlap of the TE field with the surrounding medium. Additionally, a fishbone-like SWG structure [39,40] was incorporated to reinforce the waveguide's stability and reduce the likelihood of delamination or other waveguide damage during or after fabrication. Etched openings in the oxide cladding over the sensing region allow these waveguides to be directly exposed to microfluidic flows. Three-dimensional finite difference time domain (FDTD) simulations are used to optimize the waveguide geometry and calculate the sensitivity and propagation loss (Supplementary Note 1) [39]; Fig. 2a depicts the simulated electric field intensity cross-section at two positions along the SWG waveguide, showing substantial overlap with the waveguide surface. As depicted in Figure 2a, molecular binding of biomarkers (e.g., proteins) on the surface of the SWG waveguide increases the effective index of the propagating mode, resulting in the expected red shift in the resonance wavelengths. In order to reduce resonator losses, the waveguides that connect the sensing region to the tuning and coupling regions use a combination of low-loss multimode (3 µm width) waveguides for straight routing along with single-mode strip waveguides for bends (Fig 2b). The propagation loss, caused predominantly by the surface roughness, is reduced, causing an improvement of 22% in the simulated quality factor (from 32,800 to 40,200).

The SRR's coupling region (Fig. 2c) comprises an imbalanced MZI with an FSR of ~ 2.4 nm. This scheme helps to obtain a coupling ratio that matches the resonator's losses, ensuring critical coupling. The transmission spectrum of the proposed SRR is compared with that of the MZI coupler in Fig. 2c. The modulation of the resonance and coupling interference pattern creates a characteristic transmission profile, which, when fitted, helps identify optical parameters like round-trip loss and group indices. This feature assists in characterizing resonator components using measurement data (Supplementary Equation (15)). To illustrate the transition of coupled optical power from under-coupling to over-coupling through critical coupling, the simulated transmission of the coupling MZI is shown in Fig. 2c (dotted curve).

The imbalanced coupling MZI is composed of ultrabroadband adiabatic input and output couplers designed to span the range of 1.2 − 1.7 µm while exhibiting excess loss of < 0.11 dB [41]. A thermal phase shifter was incorporated in one arm of the coupling MZI to offer additional control for adjusting the critical coupling region if required. A lightly doped rib Si-waveguide in the tuning region is used as a Thermo-Optic Phase Shifter (TOPS) to precisely sweep the resonance. The sensing region is spaced approximately 1.4 mm from the phase-shifting region to mitigate temperature impacts from the in-resonator TOPS, which could denature the sample or bioreceptor functionalization, introduce bubbles in the fluid, modulate binding dynamics, or otherwise degrade the sensing performance. We use thermal camera imaging for a chip with an activated TOPS to inspect the spatial temperature decay across the sensing axis (Fig. 2d). The measured time-varying temperature fluctuations in response to a 0.1 Hz periodic ramp heater voltage indicate the impact of thermal distancing where thermal crosstalk is reduced by 74.8 % corresponding to $\Delta T\ <\ 2°C$. This can be significantly reduced to sub milli-Kelvin when opting for more efficient phase shifters, e.g., simulated optimized TOPS with insulation that provide an order of magnitude improvement in the tuning efficiency of doped waveguide heaters, depicted in the figure for comparison[42]. With the silicon chip acting as a thermal filter, sweeping at higher frequencies would further reduce the thermal swing across the chip, effectively alleviating the impact on the sensing region (Supplementary Fig. 3c). In this work, peak interrogation was conducted at a refresh rate of 1 Hz, which is considerably higher than the rate achieved with a TLS setup (0.04 Hz for our setup).

We swept the resonance and read out the output using CMOS circuits designed and fabricated using Taiwan Semiconductor Manufacturing Company (TSMC) 65 nm technology (Fig. 2e). Ramp generation was designed using digital blocks that drive a 10-bit pulse-density modulator. Pulse-density modulation (PDM) generates effective analog voltage by adjusting the density of digital pulses. Thermal phase shifters, constrained by their heat transfer frequency bandwidth, act as low-pass filters, smoothing these pulses into a uniform analog ramp signal. The ramp frequency is controlled by tuning the division ratio, n, of the input clock signal, enabling a wide range of sweep frequencies such that $f_{ramp} = f_{PDM}/(2^{2(n+2)} \times 2^{10})$, where n={0,1,2 …7}. For an input frequency of 50 MHz, this ramp signal enables peak interrogation at frequencies of 0.18 Hz - 3 kHz. The optical transmission of the sensor is transduced to a photocurrent using an InGaAs PD connected to a 4-bit tunable passive TIA on CMOS, which provides transimpedance gain of 73 - 85 dBΩ.

Voltage level shifters were incorporated to extend the heater voltage swing to double the maximum nominal voltage tolerated by the process transistors. A 3-tap tunable termination circuit was designed with pull up/down resistance range of 35 − 325 Ω. Further details on the circuit design and the principle of operation for voltage scaling, level shifting and termination can be found in Supplementary Note 5. Using a 5V DC supply results in a resonance shift of ~ 6 FSR at a sweep rate of 1 Hz. The protein detection assays in this work were conducted at 1 Hz, resulting in resonance-scanning shifts of 4.5 FSRs.

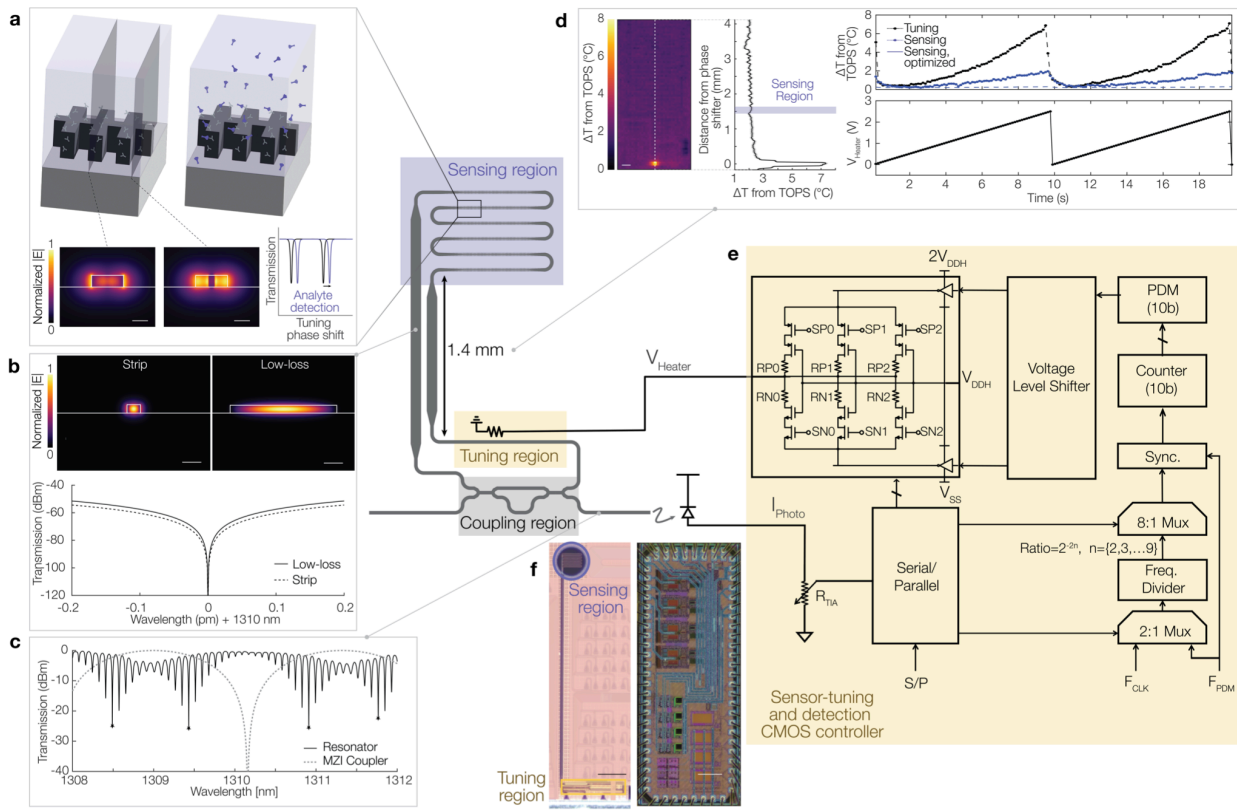

**Fig. 2 The architecture of the segmented sensor includes several distinct regions to facilitate high performance sensing.** (a) The sensing region is exposed to the sample fluid and causes the resonance peaks to shift in response to analyte binding to the waveguide surface. Sub-wavelength grating (SWG) waveguides are used in the sensing region to offer high overlap of the optical mode with the sample, yielding high sensitivity. Finite difference time domain simulations of the normalized magnitude of the electric field at cross sections through two points on the SWG waveguide show substantial overlap with the surface of the waveguide and sample, illustrating this high sensitivity (scale bar represents 0.25 μm). (b) 3 μm-wide low-loss waveguides are used for routing, improving the simulated quality factor by ~ 22% (simulated Q = 40,200 if the low-loss waveguides are used for routing, and 32,800 if regular 350 nm width strip waveguides are used for routing). To compare their mode profiles and illustrate the higher overlap of the mode with the scattering sidewall regions of the strip waveguide that contributes to these higher losses, the normalized magnitude of the electric field for both a standard strip waveguide and a low-loss waveguide are depicted with a scale bar of 0.5 μm. (c) An imbalanced MZI coupler eliminates the need to optimize coupling and facilitates coupler tuning (via an integrated phase shifter) if required. (d) The sensing and tuning regions are separated by 1.4 mm to reduce heating and temperature fluctuations in the sample fluid and sensing region. This separation reduces temperature fluctuations by >74%, from thermal camera measurements. The temperature readings versus distance shown represents the contribution of the TOPS when activated (Methods, 'Temperature characterization with thermal camera'). The mean is represented by the solid line, while the shaded region shows the standard deviation of 1960 thermal image captures. Although small periodic increases in temperature (< 1.8°C peak-to-peak) are observed in the sensing region during phase shifter modulation, these are likely significantly dampened by the presence of fluid flow at the sensor surface, and using an optimized phase shifter design with insulation is simulated to reduce these

fluctuations to 0.1℃ peak-to-peak, given that the incorporation of insulation to a doped heater scales its thermal efficiency ($\frac{\partial P_{PS}}{\partial \Phi_{PS}}$) by ~ 0.06 [42]. Scale bar for thermal image represents 0.5 mm. (e) A custom CMOS chip is used for sensor tuning and readout, incorporating functionality to both supply a pulse-density modulation (PDM) signal to sweep the tuning region phase shifter, as well as read out the resonator's transmission spectrum signal from a photodiode. (f) Micrographs of the photonic and CMOS ICs, highlighting the sensing and tuning regions on the photonic IC. The photonic and CMOS ICs have scale bars of 0.2 μm and 0.25 μm, respectively. The central schematic illustrating the resonator architecture is not drawn to scale.

## The segmented ring resonator architecture facilitates fixed-wavelength (FW) readout of resonance peak position

To illustrate that resonance interrogation provides transmission data comparable to that obtained by TLS scheme, we characterize the TOPS, create a model to relate the input sweep parameters of both FW and TLS interrogation schemes, and then compare the transmission versus both the raw voltage signal and the equivalent wavelength shift. The spectral response of the resonator is characterized by using a TLS to read out the sensor's transmission spectrum under a range of TOPS DC voltage conditions (Fig. 3a). By fitting the resonance peak shifts in the resulting spectrogram to a quadratic model (Fig. 3aii), we calculate the thermal efficiency of the implemented doped heater as $21.25\ mW/\pi$, as described in Supplementary Note 5. The PDM pulse density for a 1 Hz ramp signal, its corresponding average voltage across the TOPS, and the raw transmission signal of a segmented resonator with an air-clad sensing region over 3 ramp cycles are presented in Fig. 3b. Consistent with the transient thermal response (Fig. 2d), the transmission exhibits multiple resonance peaks with gradually decreasing intervals, attributed to the quadratic temperature (and resulting phase shift) sweep over time. Following each sweep, there is a gradual return to the initial resonance as the waveguide cools down.

With resonance shifts being linearly proportional to changes in the sensed medium ($\Delta\lambda_{res} \propto \Delta n_{clad}$), we translate our voltage axis to a linear scale. By modeling the phase shifter as a linear time-invariant (LTI) system, we linearize the acquired data, taking into account the transfer function of the TOPS and its transient thermal response (Supplementary Note 7). The transmission spectra, linearized and collected over 1000 sweeps, demonstrate good reproducibility (Fig. 3b-ii). The probability distribution function of the resonance peaks are shown for the acquired data which includes resonance drifts over time (e.g., those that may be due to temperature fluctuations or the air-clad sensing region). To further reduce the impact of amplitude noise on the measured resonance peaks, the resonance peaks in the linearized data are curve-fitted to a Lorentzian model. This gives additional precision and accuracy to the extracted resonance parameters such as the resonance peak position ($\lambda_0$), full-width at half maximum (FWHM), extinction ratio (ER) and quality factor (Q), which are later used in the peak tracking algorithm that allows us to track resonance peak shift signal during fluidic characterization of the sensor performance (Fig. 3b-iii). Gathering these data from the linearized spectrum enables using metrics of vector distance such as cosine similarity to track resonance shifts over time. After linearizing the spectra and converting the voltage to equivalent resonance shift, the data analysis algorithms used are identical to those used for simple MRR sensor data [39]. From these linearized spectra, we can compute the sensor's quality factor and compare it to that measured using a TLS interrogation scheme. Across 4 photonic chips, we measure a water-clad Q of 37,200 ± 5900 using the fixed-wavelength readout, and a similar water-clad Q of 37,900 ± 5300 using TLS. From TLS measurements, we see an improved Q prior to oxide-open and water-cladding ($Q_{oxide}$ = 56,000 ± 4000), as expected due to optical loss due to water absorption[43]. Interrogating an SRR at a specific wavelength using an FW source is thus analogous to a TLS source interrogating a fixed-coupled MRR with periodic peaks of comparable FSR and ER.

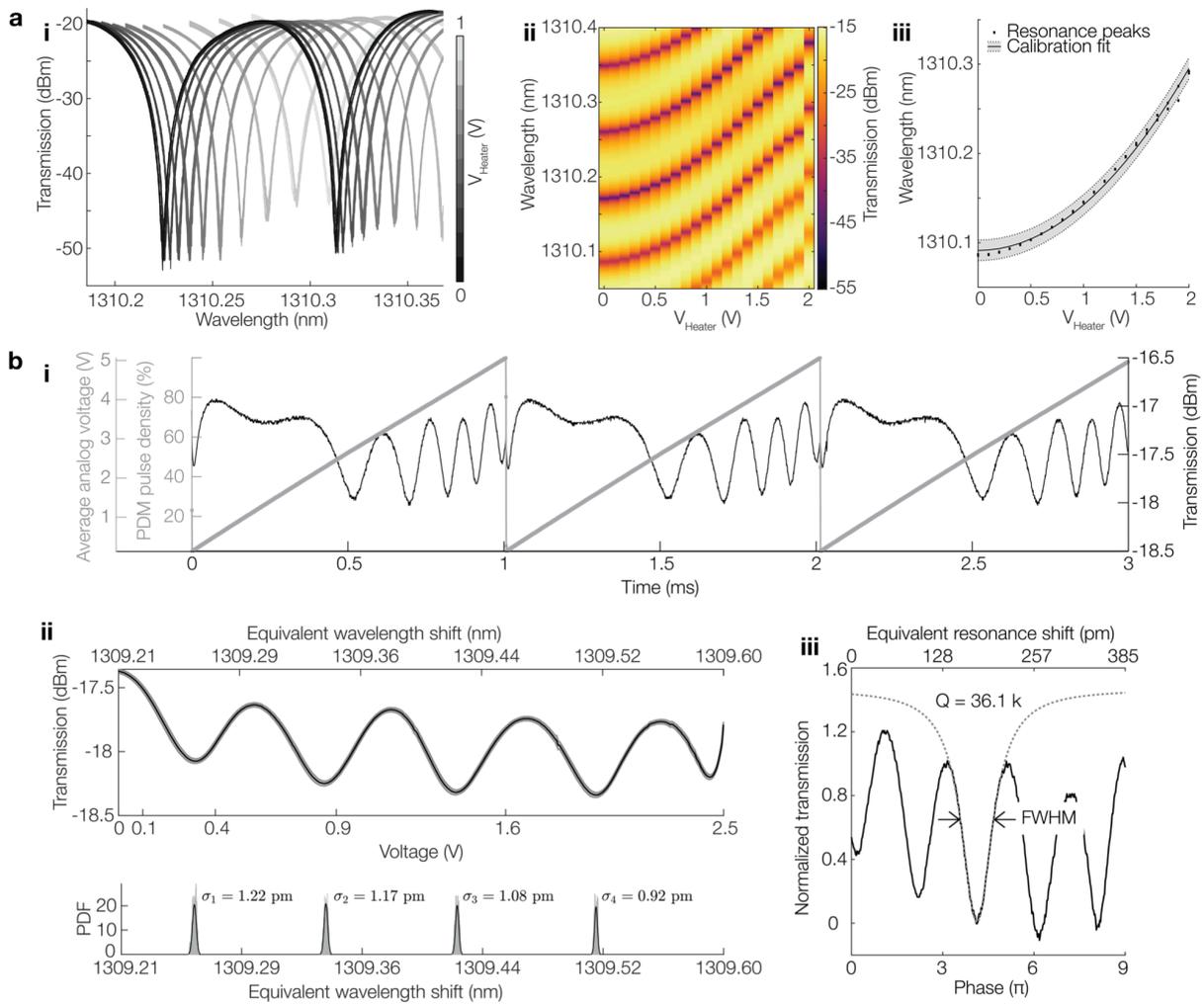

**Fig. 3 The novel segmented sensor architecture permits fixed-wavelength resonator readout.** (a) (i) Sensor transmission spectrum shifts reproducibly with phase shifter voltage, as depicted by monitoring the resonance peak position at 0.1 V steps between 0 and 1 V, with 15 spectra measurements plotted for each voltage step. (ii) A spectrogram (each column of the image depicts a reading of the resonator's transmission spectrum, measured using a TLS with optical intensity plotted on a colour scale), visualizing the resonance shift with voltage from an example voltage sweep. (iii) A quadratic fit to two replicate voltage sweeps ($\lambda = 0.05097\ V^2 + 1310.091$ nm, $R^2 = 0.9955$), used to measure the sensor's response to phase shifter voltage. The shaded region denotes the 95% confidence interval of the fit. (b) (i) The pulse density of the PDM driving signal used to modulate the sensor-tuning phase shifter, the corresponding analog voltage, and the raw transmission photocurrent waveform in response to the PDM-driven sensor tuning, depicting the raw data showing the (nonlinear with voltage) resonance peaks. (ii) Linearized readout waveform with regions of uncertainty and probability density function of the resonance peak shifts during an air-clad sensor measurement. (iii) Converting fixed-wavelength readout data to phase shift and equivalent resonance wavelength permits quantification of the resonator's quality factor at 36,100 for this example spectrum and a mean and standard deviation of 37,200 ± 5900 across 4 replicate chips.

### Our fixed-wavelength readout approach offers similar performance to tunable laser readout

To quantify the performance of the fixed wavelength laser compared to a conventional benchtop setup, we characterized the sensor's bulk refractive index (RI) sensitivity ($S_{bulk}$), stability, drift rate, and system limit of detection ($sLoD$), using both interrogation schemes (Fig. 4). We first characterized the sensor using readout with a TLS, and subsequently characterized it using the fixed-wavelength readout approach. For the TLS approach, the input wavelength was swept around the critical coupling region of the segmented resonator using a tunable laser, which is limited to a sweep rate of approximately 0.04 Hz. For the fixed-wavelength (FW) setup, we set the laser to a single wavelength around critical coupling and activated the CMOS chip to sweep the resonance at 1 Hz. FW transmission data were linearized, converted into an equivalent wavelength shift scale and then analyzed with the same approach as the tunable-wavelength data (Methods).

For both approaches, the sensor chip was integrated with a microfluidic gasket to supply fluid to the surface of the resonator, a series of salt solutions that served as refractive index standards were introduced, and the resonance peak position during a 20-min period of constant flow of 62.5 mM NaCl was used to quantify the sensor's stability and drift rate. The replicability of our segmented resonator design was assessed by conducting identical experiments on three different chips.

The spectrograms and quantified resonance peak shift vs. time plots presented in Fig. 4 (a-b) show that we can acquire similar data using the FW readout approach as in traditional passive TLS readout, at a ~98% lower system cost and a >20× higher sampling rate. By quantifying the peak shift vs. bulk RI shift data (from our RI standard solutions) and performing a linear fit for each sensor (N = 3 sensors, Fig. 4 (b-c)), we measure $S_{bulk}$ = 48.8 ± 1.3 nm/RIU for the FW readout approach (CV = 2.76%) and $S_{bulk}$ = 52.2 ± 3.6 nm/RIU (CV = 6.98%) for the TLS approach. During a 20-min stability-assessment period prior to each bulk RI ramp, we quantify drift rates of -0.7 ± 0.6 pm/min for the FW approach and -1.0 ± 0.5 pm/min for the TLS approach. After correcting the data by subtracting the drift rate, we take the standard deviation of all the measurements in the 20-min stability-assessment period to quantify the signal noise as 1.00 ± 0.30 pm for the FW approach and 0.97 ± 0.43 pm for the TLS approach. Finally, using the noise and $S_{bulk}$ data we measure $sLoD$ = 6.1 ± 1.9 × $10^{-5}$ RIU for the FW approach and 5.5 ± 2.1 × $10^{-5}$ RIU for the TLS approach. Across all metrics in Fig. 4(c), the FW sensor and TLS readout performance lies within 2 standard deviations, suggesting that performance of the two readout modalities is equivalent. Comparing our fixed-wavelength architecture with other resonators previously demonstrated in the literature, our $sLoD$ and $S_{bulk}$ present a > 5× improvement over another fixed-wavelength electronic-photonic resonator that employed intensity interrogation, which presented $S_{bulk}$ = 1.95 nm/RIU and $sLoD$ = 3.5 × $10^{-4}$ RIU [34].

The results presented in Fig. 4 show that the FW sensor readout approach does not appear to affect sensor performance using our metrics of interest; however, the use of the segmented sensor architecture is expected to reduce the resonator's sensitivity compared to a simple ring resonator with the entire device immersed in fluid, since only a portion of the resonator length is exposed for sensing (Supplementary Equation (11) and Supplementary Note 2). Indeed, our measured $S_{bulk}$ of ~50 nm/RIU is ~14% of the simulated value for the same fishbone SWG waveguide architecture (357 nm/RIU) and ~15% of the measured value for a simple ring using this waveguide architecture (332 nm/RIU) [39]. This reduction in $S_{bulk}$ is in line with prediction using our ~18.8% fill factor, with a slightly lowered sensitivity compared to prediction owing to the ~30 nm of oxide cladding remaining at the base of the 220 nm waveguides after oxide-open. Nevertheless, the intrinsic limit of detection calculated from our measured Q and $S_{bulk}$ using FW readout ($iLoD$ = 7.4 × $10^{-4}$ RIU) remains competitive with the simple ring (iLoD = 5.8 ± 0.9 × $10^{-4}$ RIU [39]), owing to the improved quality factor afforded by the segmented sensor architecture. Although our sensor's performance does not yet match that of previously demonstrated simple strip waveguide resonators using tunable laser readout ($S_{bulk}$ = 163 nm/RIU and $sLoD$ = 7.6 × $10^{-7}$ RIU)[17], the performance of our sensor can likely be further improved by employing additional averaging as well as reference sensors to compensate for thermal fluctuations. There also remains potential for further improvement over this first-generation architecture by optimizing the sensor design to improve the fill factor, for example by employing additional thermal isolation techniques, alternate phase shifter designs, or increased length in the sensing region. Overall, the results of Fig. 4 show that our fixed-wavelength readout and segmented architecture outperform previous fixed-wavelength resonator designs and exhibit performance suitable for biomolecule sensing.

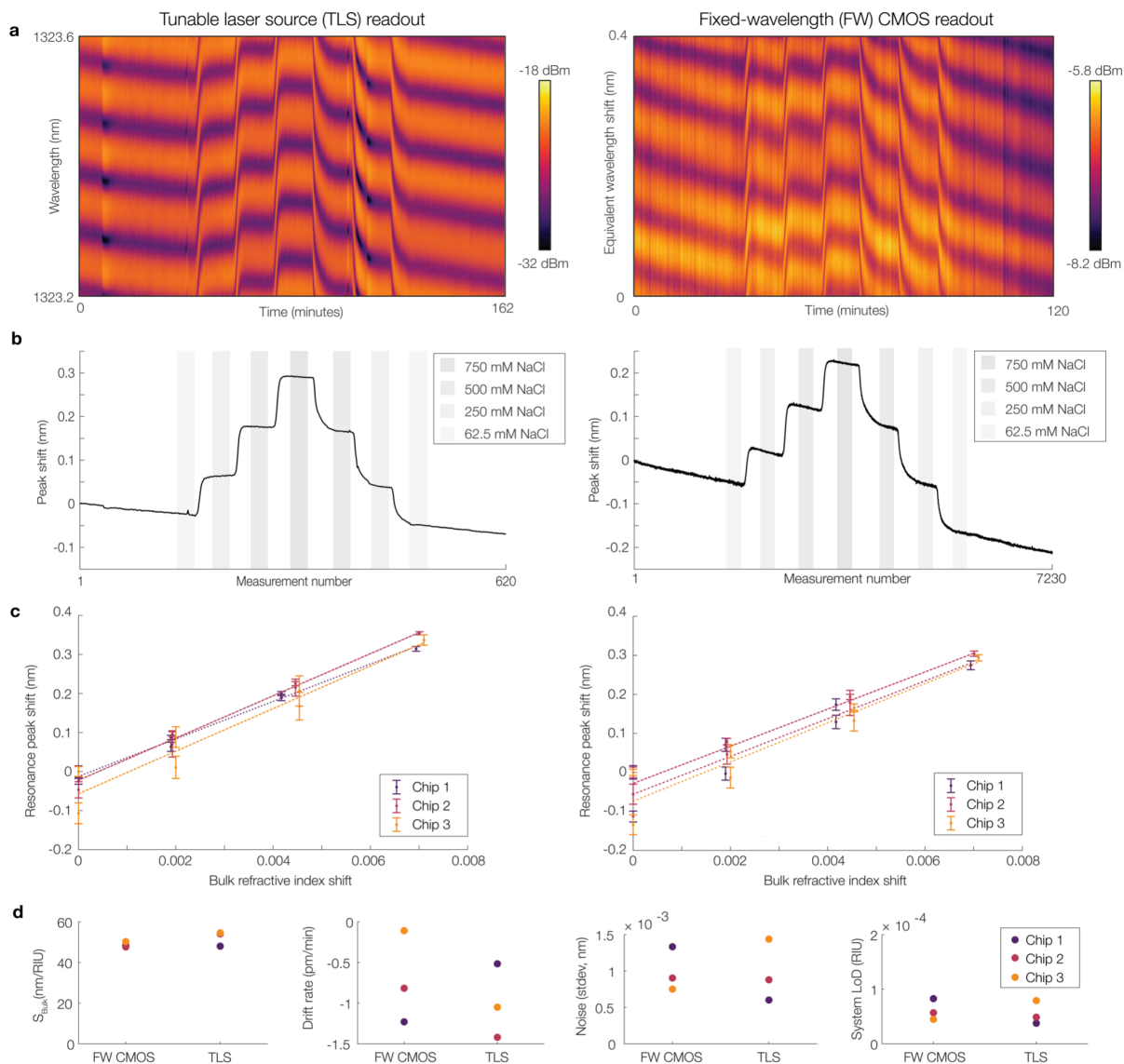

**Fig. 4 Fixed-wavelength (FW) sensor readout (right) offers comparable sensor performance to TLS readout (left) at ~98% lower cost.** (a) Spectrograms comparing the raw resonator signal during a sensor characterization experiment using swept-tunable and fixed-wavelength laser readout. (b) Resonance peak shift plots quantified from the same data as in (a), showing overlays of the regions that were quantified to measure peak position in each refractive index standard solution (shaded boxes). (c) Quantified peak shift vs. change in standard solution bulk refractive index, for 3 replicate experiments. Datapoints depict the average resonance peak shift compared to the 62.5 mM solution, and error bars depict the standard deviation of N = 25 (for the TLS readout) and N = 250 (for the FW readout) measurements in each refractive index standard solution. Dashed lines depict the curve fits used to extract the bulk refractive index sensitivity for each sensor. (d) Comparison of bulk RI sensitivity $S_{bulk}$, drift rate, resonance stability noise, and system limit of detection for TLS and fixed-wavelength readout of each sensor, showing similar performance for the two readout approaches. Full-system cost comparison is described in Supplementary Table 1.

## The novel segmented microring resonator architecture facilitates biomolecular detection with fixed-wavelength resonance peak-tracking readout

After characterizing and comparing the performance of our fixed-wavelength readout and segmented sensor architecture using refractive index standard solutions (Fig. 4), we sought to demonstrate its ability for biomolecule detection using a demonstration binding assay (Fig. 5). Using the same microfluidic device and automated fluid control system as was used to supply fluids to the surface of the resonators for the data in Fig. 4, we performed a multi-stage binding assay to detect the SARS-CoV-2 spike protein. We used microfluidic flow-mediated functionalization with antibodies using a bioaffinity-based functionalization technique employing protein A, which binds the Fc region of various antibody subtypes to facilitate potentially oriented immobilization of antibodies on the resonator surface[22]. This assay type allowed us to quantify resonance peak shifts during each stage of functionalization (protein A and antibody binding) and detection (spike protein detection), as well as quantify the nonspecific signal during sensor blocking with bovine serum albumin (BSA).

Fig. 5(a-b) depicts the stages of the binding assay as well as the resonance peak shift data acquired using FW readout during each assay stage. Clear peak shift signal is visible during the protein A, antibody, and spike protein assay stages, with small shifts due to nonspecific binding of BSA. Although the full assay duration was ~200 minutes, encompassing stability characterization at the start and end of the assay as well as functionalization and detection, the spike protein detection itself lasted only 17 minutes. This duration, which is governed by the binding kinetics of the antibody-antigen pair and the microfluidic transport of the biomolecules to the surface of the resonator[25], is similar to that for simple ring resonator sensors and could be further improved by scrutinizing the initial slope of the resonance shift dynamics rather than the total shift, as has been previously used for analysis of silicon photonic[28] and SPR [44] sensor data.

Quantifying the peak shift data for the 3 replicate trials (Fig. 5(c)) reveals peak shift signals of 0.205 ± 0.019 nm (Coefficient of Variation (CV) = 9.4%) for the protein A adsorption, 0.27 ± 0.07 nm (CV = 26%) for the antibody immobilization, 0.19 ± 0.03 nm (CV = 15%) for the spike protein detection, and a small signal (0.03 ± 0.05 nm) in response to the nonspecific binding/blocking stage of the assay. The inter-assay CV for spike protein detection is in line with those previously demonstrated in the literature for other silicon photonic sensors, as well as close to values reported for commercial ELISA kits. Although one might expect that a larger shift from adsorbed antibody during functionalization (potentially indicating that more specific binding sites are present on the resonator) might correlate with larger spike protein detection shifts, we do not observe any obvious correlation between the peak shift signal for the antibody immobilization and spike protein detection stages. This may indicate the presence of antibody aggregates or other factors that increase the antibody-binding signal without increasing the available binding sites for spike protein. Nevertheless, we observe clear, detectable signal in all trials, showing the utility of our fixed-wavelength readout and segmented sensor architecture for biomolecule detection.

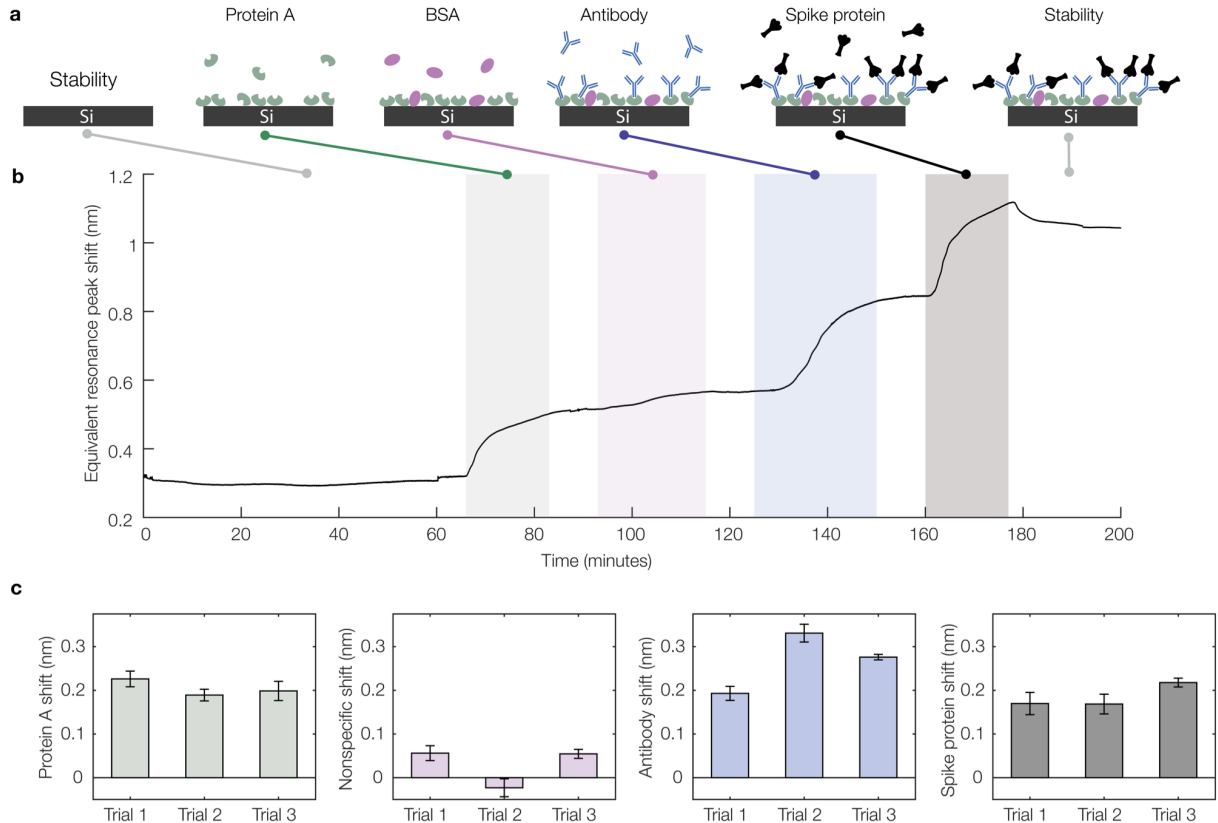

**Fig. 5 Demonstration binding assay using the segmented sensor architecture and fixed-wavelength readout, showing detection of the SARS-CoV-2 spike protein.** (a) Schematic depicting the steps of microfluidic waveguide functionalization and binding assay detection of SARS-CoV-2 spike protein in buffer solution. (b) Sensor resonance peak shifts for a representative binding assay, acquired using fixed-wavelength sensor readout and showing the peak shift dynamics during each stage of the functionalization and detection. Regions without coloured overlay boxes depict PBS buffer flow washes. Resonance peak shift data were corrected for drift in PBS using a linear fit to an initial stability region in PBS buffer at the beginning of the experiment. (c) Quantified peak shifts for each stage of functionalization and detection (shift after protein A deposition to facilitate antibody binding, quantified nonspecific shift after blocking with bovine serum albumin (BSA), shift from antibody immobilization to functionalize the surface of the resonator, and shift from spike protein detection) on N = 3 replicate sensor chips. The quantified data for each trial depicts the mean shift and propagated error (standard deviation) from calculating the average and standard deviation of the resonance peak positions from 150 measurements during PBS buffer washes before and after each assay stage.

## Discussion

In summary, our novel segmented ring resonator architecture addresses the complexity and high cost of conventional integrated photonic resonator-based biosensor setups while maintaining the advantages afforded by resonance-tracking interrogation. Rather than tuning the input wavelength, we use a fixed-wavelength laser to tune the resonance via a thermo-optic phase shifter, with the undesired thermal effects mitigated by distancing the biosensing region from the heating region. Our design uses compound routing to minimize the propagation loss due to scattering. Our fixed-wavelength resonance-tracking interrogation scheme showed similar performance as TLS resonance-tracking interrogation across several metrics, and we also demonstrated that the SRR sensor architecture is suitable for replicable biomarker detection using a binding assay to detect the SARS-CoV-2 spike protein. Future work towards using this new architecture to address open needs in decentralized diagnostics will include sensor multiplexing and further miniaturization of the readout system to chip-scale, as well as assay development and optimization to detect biomarker targets of interest.

Online Methods

**PIC Design, Simulation and Fabrication**

Sub-wavelength grating (SWG) waveguides were designed with a width of 0.5 µm, a period of 0.2 µm and a duty cycle of 0.5. A 0.1 µm fishbone waveguide was used to enforce the structure against potential delamination during the oxide-open process or failure during surface functionalization or drying, as described in our previous work [40, 39]. Within the resonator structure, we used multimode 3 µm waveguides to route the light between the sensing and tuning regions to significantly reduce scattering loss, while 0.35 µm strip waveguides were used for all bends to avoid exciting higher order modes. Adiabatic Bezier curves with Bezier number of 0.25 were used to reduce bending loss [45]. Splitters and combiners were implemented using ultra-broadband adiabatic SWG-assisted couplers of low excess loss and wide adiabatic coupling spectral range[41].

To accommodate sensor tuning, an in-resonator TOPS was used [46]. In our implementation, the TOPS consists of a 150 µm long lightly doped rib waveguide. The waveguide's core width is 0.45 µm, with a slab width of 7.35 µm and a slab thickness of 90 nm. To form ohmic contacts, highly doped regions, each 1.5 µm wide, were added on both sides of the rib waveguide. This creates a low resistive element across the rib waveguide that can be used for phase shifting. In addition, the in-coupler TOPS, designed to offer additional tunability of the coupling, also comprises a 150 µm long rib waveguide with core width of 0.45 µm, a slab width of 7.35 µm and a slab thickness of 90 nm, with 0.5 µm-wide highly doped regions on either side of the rib waveguide. These highly doped regions each have six metal contacts, resulting in voltage being supplied across five resistive elements in parallel, along the length of the waveguide. While the in-coupler TOPS adds additional tunability to match resonator response to a given fixed wavelength and will be important to align the transmission spectra in multiplexed resonators interrogated with a single fixed-wavelength source, it was not used in the experiments reported here.

The strip and rib waveguides' effective index ($n_{eff}$), group index ($n_g$), susceptibility, and loss for strip and rib waveguides were simulated using Ansys Lumerical MODE Optical Waveguide & Coupler Solver. For the SWG waveguide, finite-difference time-domain band-structure simulations were performed using Ansys Lumerical FDTD for a single unit of the waveguide with Bloch boundary conditions to extract the wavevector, from which the effective and group indices were calculated. The exposed SWG waveguide was designed in a cutback structure with a length of 1 mm, constituting a fill factor of 18.8% of the resonator's 5.3 mm total length.

The photonic circuit design was fabricated using the AIM Photonics Base Active PIC process on an SOI wafer with 220 nm thick silicon waveguides and a 2 µm layer of buried oxide. Individual diced PIC chips were 2.066 mm (height) × 4.35 mm (width), with the sensing region located approximately 1 mm and 1.65 mm from the left and bottom edges of the chip, respectively. Oxide cladding is deposited as a part of the active PIC fabrication process, and we developed an in-house post-fabrication process to remove most of the $SiO_2$ cladding in a circular region over the SRR's sensing region to expose the sensing region waveguides to fluid. A nitride layer was incorporated around the sensing region at the design phase to serve as an etch-stop layer and facilitate the oxide opening process. The PIC was first cleaned by sonicating in acetone, rinsing in isopropanol (IPA), and drying with nitrogen ($N_2$). A 30-s $O_2$ plasma treatment was used to clean the surface, and the chip was baked at 110 °C for > 5 mins to dehydrate the surface and prepare for photoresist deposition. AZP4620 photoresist was then applied and spun at 4500 RPM for 1 min at an acceleration of 500 RPM/s. After spin-coating, the chip underwent a post-exposure bake at 110 °C for 2 min. An exposure dose of 1,000 mJ/cm$^2$ was delivered using a maskless lithography system (Heidelberg Instruments MLA 150 Maskless Aligner), and exposed photoresist was subsequently removed using AZ MIF300 developer over 4 min, followed by rinsing well with deionized (DI) water and drying using $N_2$.

Deep reactive-ion etching (DRIE) was performed using an SPTS Rapier DRIE tool located in the 4D Labs nanofabrication facility, utilizing gas flow rates of 65 sccm for $C_4F_8$, 35 sccm for $O_2$, and 300 sccm for Ar, with respective durations of 6, 6, and 1.5 minutes. The surface was then rendered hydrophilic using a 30 s $O_2$ plasma treatment. For the 3 chips reported in Figures 4-5, the measured residual cladding oxide thicknesses were 30 nm for Chip 1, 27 nm for Chip 2, and

36 nm for Chip 3. The oxide-open process was finished by wet etching in buffered oxide etchant (BOE) 10-1 (UN2187 ammonium hydrogen difluoride solution, 8-6.1, PGII, Transene Company Inc., USA), periodically checking residual oxide thickness using a contact profilometer (Dektak XT, Bruker, USA) and a reflectometry-based thin-film thickness measurement system (Filmetrics F20, USA). The wet etching rate was estimated as 1 nm/s, and an initial etch time of 4 minutes was used. An iterative etching-measurement process using this estimated etch rate was then used to gradually reach the desired etching depth. We aimed for ~30 nm remaining $SiO_2$ cladding thickness at the bottom of the waveguides to avoid waveguide undercuts that might be incompletely wetted and serve as bubble nucleation sites during microfluidic flow [47].

**Microfluidic device fabrication**

A 2-layer microfluidic gasket was custom-designed to deliver fluid to the surface of the SRR's sensing region and fabricated out of poly(dimethylsiloxane) (PDMS) using methods adapted from those previously reported[39,48,49]. The sensing region of the SRR has an area of approximately 150 µm × 100 µm, and the layer 1 microfluidic channel with rectangular cross-section was designed with a width of 250 µm, height of 200 µm, and a length of approximately 0.9 mm in contact with the surface of the PIC in order to fully cover the sensing region with generous alignment tolerance. Square vias (250 µm × 250 µm × 1 mm height) located at each end of the layer 1 channel routed the fluid between the layer 1 and layer 2 channels. The layer 2 channels routed the fluid between layers of the elastomeric PDMS material, between the vias and the fluidic input/outputs (I/Os). The total microfluidic device thickness was 4 mm, with the layer 1 piece (housing the layer 1 channel, through-holes for the fluidic vias, and through-holes for alignment bolts) having a thickness of 1 mm and the layer 2 piece (housing the layer 2 channels, fluidic I/O through-holes, and through-holes for alignment bolts) having a thickness of 3 mm.

The fluidic channel layers, along with alignment marks and markings to guide a manual cut of the gasket edge to accommodate the electrical wire bonds and fibre array, were designed along with the PIC layout using KLayout software, and exported as .dxf files for each layer to be extruded. The .dxf files were imported into SolidWorks software (Dassault Systèmes) and extruded into 3D moulds for the two microfluidic gasket layers. These two 3D moulds were 3D-printed in-house using a 3D printer and resin optimized for microfluidic mould fabrication (PROFLUIDICS 285D printer and Master Mold resin, CADworks3D). After printing was complete, the moulds were removed from the build plate and postprocessed with two isopropanol washes (5-20 mins duration) and an ultraviolet cure (~40-60 minutes from the top side and ~20-40 minutes from the backside using the CADWorks3D Cure Zone curing unit). After postprocessing, the moulds were cast once with PDMS and cured at 60 °C in an oven overnight. This first cast of PDMS was discarded, as it tends to contain residual uncured resin residue. After the first cast, the moulds are ready to be used for multilayer gasket fabrication.

Multilayer gaskets were fabricated using a soft lithography[50] and an on-ratio partial-cure PDMS diffusion-bonding protocol[51]. A 10:1 mixture of Dow Sylgard™ 184 PDMS base and curing agent (Ellsworth Adhesives) was mixed using a planetary and centrifugal mixer (THINKY USA), poured into each mould such that the mould was slightly overfilled, and degassed for 30-60 mins in a vacuum desiccator (Bel-Art™ Space Saver, Fisher Scientific) until no bubbles remained. As previously described[48], in order to create flat gaskets with through-hole vias, a sheet of transparency film (Canon 6101AJ28AA Transparency Type E) was slowly lowered onto the mould surface. A flat rectangular piece slightly larger than each mould (laser-cut from 3 mm-thick acrylic) and ~500 g weight were then placed on top of the transparency for each mould to provide even pressure during curing and help define the through-holes. The two filled moulds were cured at 60 °C for 80 minutes, and then the moulds were removed from the oven and transparency films peeled off, PDMS carefully demolded. The PDMS pieces were checked to ensure that their through-hole vias were clear, and the two layers were subsequently aligned with the assistance of a 3D-printed alignment jig and placed back into the 60 °C oven overnight with weights to complete curing and bonding.

**PIC chip preparation and printed circuit board (PCB) assembly**

An electroless nickel immersion gold (ENIG)-coated carrier PCB was custom designed with a thermal conductive pad matching the size of the chip and vias to dissipate heat to the lower side of the PCB. Screw holes were included in the PCB to facilitate assembly and alignment of the PIC, thermal control Peltier element, heat sinks, and microfluidic gasket. A thermally conductive adhesive epoxy (MG Chemicals P/N: 8349TFM-25ML, Mouser P/N: 590-8349TFM-25ML) was used to attach the oxide-opened PIC to the conductive pad. A 0.8 mm-thick PCB shim was designed with a slot to ensure accurate alignment of the PIC on the thermally conductive pad and secure it during the epoxy curing process.

The electrical bond pads on the PIC were connected to the header PCB using 1-mil thick aluminum bond wires using a West-BondManual Wedge Bonder (Model 747630E-79). We subsequently affixed the PCB to a thermal control and microfluidics stack. From bottom to top, the stack consisted of: (1) commercial heat sinks affixed with thermal paste (Noctua NT-H2); (2) a 3.5 mm-thick custom-machined aluminum mounting plate with #4-40-tapped bolt holes; (3) a commercial thermoelectric cooler module (Mouser 490-CP60133, 15 × 15 × 3.3 $mm^3$); (4) a 3.5 mm-thick custom-machined aluminum thermal plate with through-holes for the bolts and a 3 mm-deep inset to house a 10k negative temperature coefficient (NTC) thermistor (Vishay THERMISTOR NTC 10KOHM, PN:NTCLE100E3103JB0) and its leads (the groove was filled with thermally conductive epoxy to stabilize the thermistor); (5) carrier PCB with affixed PIC; (6) 0.8 mm-thick shim PCB surrounding the PIC (slightly less thick than the PIC) to help support the microfluidic gasket; (7) a custom multilayer microfluidic gasket fabricated using PDMS and reversibly bonded to the PIC; (8) a custom-fabricated washer, laser-cut out of rigid 3 mm-thick acrylic, to help provide even pressure to secure the elastomeric microfluidic gasket. The stack was aligned and secured using #4-40 nylon threaded rods (McMaster-Carr) and thumb nuts (McMaster-Carr). The thermal control setup senses and conducts temperature from the thermal pad to the Peltier module. Thermal paste (Noctua NT-H2) was applied between the PCB, aluminum thermal plate, Peltier module, mounting plate, and heat sinks to improve thermal conduction between the layers.

For this initial proof-of-principle work, an off-chip laser was used to provide the fixed-wavelength input light, and a polarization-maintaining 4-channel optical fiber array (FA, OZ Optics LTD., VGA-4-127-8-A-6.9-2.5-1.23-P-1550-8/125-3A-1-2-0.5-GL-NoLid-Horizontal) was used to couple light from the laser to the chip. To mitigate mechanical noise or optical attenuation from mechanical drift, the FA was securely attached to the PIC header PCB. We used UV resin (Limino UV Resin Clear) to attach the FA to a ~15 mm (height) × 25 mm (width) × 1 mm (thick) glass shim which was then affixed to the 0.8 mm shim PCB with thermal epoxy. A motorized optical stage (Maple Leaf Photonics) was used to assist in alignment and correction during the curing of the thermal epoxy.

**Sensor readout setup**

Light from a tunable O-band laser (Agilent 81672B) was coupled to the photonic chip through a polarization-maintaining (PM) fiber array. Depending on the type of testing (tunable laser (TLS) vs. fixed-wavelength readout), the laser provided either swept-tunable or fixed wavelength light input. The input light is split on-chip between a segmented ring resonator and a loopback grating coupler, which was used for fine alignment and input laser intensity monitoring. For the TLS readout setup, mainframe detectors (Agilent 81635A) were used to detect the output light from the loopback as well as from the SRR. A custom Python software GUI was used to control the laser and detector, save each optical spectrum, and record measured chip temperature and timing for each sweep for subsequent analysis.

In our fixed-wavelength biosensor readout setup, we use a fiber-coupled InGaAs PIN photodiodes (PM-Optics, SMF-28 fiber, FC/APC connector, DPIN-23133) for optical transduction. Analog Discovery 2 (AD2) was selected for its compactness and portability, serving as a USB oscilloscope to digitize, serialize and log data. The CMOS chip holding the driver and TIA designs was wire-bonded to a CQFP 80 packaging and soldered to a custom-designed carrier PCB. A main PCB was designed to interface between the EIC, PIC, InGaAs PD and AD2 as pluggable modules. The temperature was stabilized at 22 °C for both readout setups using a TEC controller (LDC501). The PID parameters of the controller were auto-tuned to maintain a temperature standard deviation of < 0.5 mK over the course of 14 hours.

## Temperature characterization with thermal camera

The photonic chip, mounted with the stack shown in Supplementary Fig. 3, was placed under the camera (VarioCam HiRes 1.2 M, Jenoptik AG, Germany) and brought into focus. An arbitrary waveform generator (RIGOL DG1022) was used to drive the TOPS, and a TEC controller (SRS LDC501) was connected to the in-plate NTC thermistor and thermoelectric cooler described above, with the temperature set to 22°C. Data were acquired for various combinations of frequency and voltage range driving the heaters, as detailed in Supplementary Note 3.

To establish a temperature baseline, readings were taken with the TOPS deactivated (set to 0 V). These ambient readings were subtracted from measurements acquired with the heater activated in order to isolate and analyze the heating contributed by the activated TOPS. The temperature gradient data across the chip, shown in Fig. 2d, were collected in two runs, each with 1960 thermal image captures taken over a period of 2 minutes. The first run had the TOPS deactivated to obtain the chip's temperature at ambient conditions. In the second run, the TOPS voltage was set to 2.5 V (DC). To extract the temperature difference contributed by the TOPS, we subtracted the temperature obtained in the second run from those obtained with TOPS deactivated. The pixel resolution of the thermal imaging camera used to measure the temperature profile of our chip is 16 µm × 16 µm. Given the different sizes of TOPS (~150 × 7.35 µm$^2$) and the biosensing region (~130 × 90 µm$^2$), the temperature readings are averaged over different areas such that a 6 × 1 pixel window was used to cover 90 × 15 µm$^2$ of the TOPS and a 3 × 3 pixel window to cover 45 × 45 µm$^2$ sensing region, respectively. For characterizing the spatial temperature gradient across the PIC, a DC voltage of 2.5 V was applied to the TOPS. The transient response was acquired for TOPS activated with a 0.1 Hz sawtooth wave, 2.5 V peak-to-peak. Thermal images were captured at 25 fps and processed using custom MATLAB scripts. An image processing script was developed to quantify temperature and time from the temperature images, which represented temperature with a colormap and included a colorbar on each image to define the plotted temperature range. We generated a look-up table (LUT) of RGB colour values and the temperatures they represent from the colorbar and its temperature limits. The script incorporated optical character recognition (OCR) to quantify the time and minimum/maximum temperatures for each image's colorbar. To analyze transient temperature readings at specific chip locations, the pixels within the TOPS and sensing region regions of interest were quantified and their RGB values were translated into temperature readings. This is achieved by constructing an array of RGB values from the temperature colorbar and computing the Euclidean distance, $d$, between the pixel RGB readings and the colorbar array entries, such that $d = \sqrt{\sum_{i=1}^{3}(Image_{pixel,i} - Colorbar_{pixel,i})^2}$, where $Image_{pixel}$ and $Colorbar_{pixel}$ are vectors in the RGB Euclidean space. Using the image colorbar vector as an LUT, we interpolate the temperature value that corresponds to the colorbar with the minimum Euclidean distance.

## Microfluidic testing system

After fabricating and integrating the 2-layer PDMS microfluidic gasket with the PIC and thermal control stack, a pressure-driven automated fluid control system (Fluigent® LineUP™) was used to deliver an automated sequence of fluids to the surface of the sensor for our fluidic testing protocols. Our system comprises an electronic pressure regulator (Fluigent® Flow EZ™) that pressurizes the headspace of fluidic reservoirs in order to drive fluid flow, a 10-channel switch that selects between 10 pressurized reservoirs to flow fluid (Fluigent® M-SWITCH™), bidirectional calorimetric flow sensor to provide feedback for the flow rate control algorithm (Fluigent® Flow Unit M), and inline passive bubble trap with hydrophobic membrane (PreciGenome EZMount with 25 µL internal volume) to help remove bubbles from the flowing fluid. The different components of the fluid control system are connected using fluorinated ethylene-propylene (FEP) and polyether ether ketone (PEEK) tubing. A piece of the same 1/32" OD × 0.010" ID PEEK tubing that was used for the outlet from the inline bubble trap (Idex 1581L, Cole-Parmer Canada) was friction-fit directly into the 0.5 mm diameter circular I/Os in the PDMS gasket to serve as the fluidic inlet. Another ~10 mm-long piece of the same PEEK tubing was connected to a ~30 cm length of 1/16" OD, 0.02" ID Tygon® tubing (Masterflex® Microbore ND-100-80, purchased through Cole-Parmer Canada but no longer available) and friction-fit into the gasket output hole to serve as the fluidic outlet.

In order to reduce the likelihood of bubble nucleation in the microfluidic device during long experiments, we performed initial slow manual prewetting of the microfluidic gasket with anhydrous ethanol (P016EAAN, lot 035057), which is expected to more effectively wet small crevices in the device that can serve as bubble nucleation sites[47], as compared to water (due to the lower contact angle of ethanol on the PDMS and PIC materials). A syringe and manual delivery were used for this initial prewetting because the bubble traps in the fluid control system are not compatible with organic solvents like ethanol. To facilitate this initial ethanol wetting, along with rinsing with water and subsequent connection to the automated fluid control system without introducing bubbles into the microfluidic system, we used a short ~3 cm length of chemically resistant Tygon® 2375 tubing (VWR 89178-230) along with hose-barb fittings (Cole-Parmer RK-02023-82) and a Luer-lock adapter (Cole-Parmer RK-02014-15 and RK-02007-00) alongside FEP tubing sleeves to interface the 1/32" PEEK tubing to the connectors (Cole-Parmer RK-02018-08). Using this system, we slowly delivered ~1 mL anhydrous ethanol to the microfluidic channel through a 3 mL syringe, Luer adapter, FEP tubing, hose barb fitting to Tygon® 2375 length, and finally the hose barb fitting to the PEEK tubing connected to the gasket inlet. We then blocked the tubing outlet Tygon® tubing using a ratchet-style pinch-clamp to stop the flow and carefully disconnected the ethanol-filled syringe while supplying gentle pressure to the plunger to fill the Luer adapter and avoid introducing bubbles into the system. Another syringe filled with ultrapure water (ASTM Type I water from a Nanopure Diamond system) was subsequently connected, outlet unblocked, and the microfluidic device was rinsed with ~2 mL ultrapure water in order to avoid potential precipitation of salt within the channels in case residual salt in the fluid control system made contact with the ethanol. The outlet was then blocked again, and the hose-barb fitting connected to the syringe was slowly disconnected from the Tygon® 2375 tubing and the hose barb fitting connected to the automated fluid control system was connected to the Tygon® 2375 tubing while providing a constant-pressure 50 mbar fluid delivery of ultrapure water. Finally, the outlet was unblocked and the fluid control system was used to provide constant-flow rate delivery of water to further rinse the system while we set up the fluidic delivery protocol and acquisition of sensor data. All fluidic testing protocols were developed in the Fluigent OxyGEN software and ran automatically. All reservoirs to be delivered during a testing protocol were connected to the fluid control system and the lines for each reservoir primed with fluid prior to gasket prewetting and fluid control system connection, in order to prevent delivery of air from empty fluid lines to the gasket.

**Sensor data analysis**

To analyze the sensor data, we first correct for the nonlinearities introduced by the quadratic behavior of the TOPS and the signal smoothing caused by the slow thermal dissipation. We model the thermal dissipation of our chip mount stack with an RC circuit analogy, a widely used technique for characterizing thermal behavior in power modules using lumped thermal resistances and capacitances (Supplementary Figure 3) [52,53].

Our approach begins with obtaining a stream of raw SRR transmission data, which are then divided into individual pre-spectra (Supplementary Fig. 7). The thermal model is then applied to linearize the data. In the resulting linearized spectra, we fit the resonance peaks to Lorentzian models. Finally, we track the resonance peaks by matching sets of peaks across consecutive sweeps using a cosine similarity cost function based upon weighted vectors of Lorentzian fit parameters for each peak (Supplementary Note 7). Through this approach, we generate sensorgram data reporting effective resonance peak shift vs. time, and subsequently analyze these sensorgram data with approaches that depend on the testing type, as described in the sections below.

**Resonator performance characterization testing**

To characterize the resonators' bulk refractive index (RI) sensitivity ($S_{bulk}$), stability, drift rate, and system limit of detection ($sLoD$), the sensors were exposed to a sequence of sodium chloride (NaCl) solutions that served as bulk refractive index standards. A 2.5 M NaCl stock solution was prepared using ultrapure water and NaCl solid (Fisher S271-3, lot 166221), and this stock solution was diluted to prepare the bulk refractive index standard solutions with the concentrations of interest (62.5 mM, 250 mM, 500 mM, 750 mM). The refractive indices of the standard solutions were measured using a refractometer (Spectronic Instruments Abbe-3L), immediately prior to each experiment.

After microfluidic setup, fluid control system priming, and prewetting, an automated protocol was delivered to supply a controlled sequence of fluids to the sensor at a constant flow rate of 30 µL/min. This protocol included a sequence of 15-min steps of increasing and decreasing bulk refractive index (62.5 mM NaCl, 250 mM NaCl, 500 mM NaCl, 750 mM NaCl, 500 mM NaCl, 250 mM NaCl, 62.5 mM NaCl), as well as 30-min stability periods (where ultrapure water or 62.5 mM NaCl were delivered under constant flow) both before and after the bulk refractive index standard steps. Both upwards and downwards ramps in the refractive index were performed to check for hysteresis in the system and average out any sensor drift due to slow waveguide etching or other factors. To facilitate comparison, we ran each performance characterization protocol twice on each sensor: once with TLS readout, and then once with fixed-wavelength readout. In both protocols, the laser or resonator was continuously swept to facilitate continuous readout; however, the readout frequency was faster for fixed-wavelength readout (1 Hz) compared to TLS (0.04 Hz) due to hardware limitations. The ambient temperature of the thermal stack was measured with the in-thermal-plate thermistor as ~25 °C, and the thermal control temperature was set just below this, to 22 °C, for all fluidic testing.

After linearizing the fixed-wavelength readout data and performing resonance peak fitting and tracking to extract the effective resonance peak shift as a function of time as described above, a custom semi-automated MATLAB® script was used to compute the resonator performance metrics from the tracked peak-shift data. This script took in a list of solutions delivered and their bulk refractive indices, and allowed the user to click on the regions of the effective peak shift vs. time plot for each bulk RI standard. For each region, a set of 25 (for TLS) or 250 (for fixed-wavelength) peak position data points were averaged to obtain the average resonance peak position in each standard solution. These average resonance peak position vs. measured bulk RI data were then least-squares fitted to a line function to extract the bulk refractive index sensitivity (slope of the line) and goodness-of-fit data. The MATLAB® script also allowed the user to click on stability regions of interest, and we analyzed 20-minute regions of the peak shift data to extract the resonator performance. The resonator stability was analyzed in multiple ways. First, the raw noise was calculated as the standard deviation of the resonance peak position during the full 20-minute stability region. Second, the drift rate was computed from the slope of a linear least-squares fit to the data in the stability region. Third, the noise after drift correction was computed as the standard deviation of the raw stability data after subtracting the linear fit (compensating for sensor drift, which can be accomplished in a sensing experiment by including a stability characterization period at the beginning of the assay and correcting the peak shift data). The system limit of detection ($sLoD$) was computed as three times the noise standard deviation divided by the bulk sensitivity ($sLoD = 3\frac{\sigma}{S_{bulk}}$). Both the raw and the drift-corrected $sLoD$s were computed (from the raw and drift-corrected noise standard deviations, respectively), and the drift-corrected noise σ and $sLoD$ are presented in Figure 4.

**Demonstration binding assay to detect SARS-CoV-2 spike protein**

After characterizing the sensors' performance using the fluidic testing protocol described above, we moved to running the demonstration binding assay without removing the gasket or drying the channel. The demonstration binding assay delivered a sequence of reagents to the surface of the sensor to first functionalize the sensor for specific detection of SARS-CoV-2 spike protein using bioaffinity-mediated functionalization, and subsequently specifically detect spike protein.

We first exposed the sensor to a 60-min flow of phosphate-buffered saline solution (PBS, Gibco™ 10010049) to characterize the baseline stability and facilitate baseline-correction. A 100 µg/mL solution of protein A (Sigma-Aldrich P6031, lot 0000142821), a protein that specifically binds the Fc region of certain antibodies to facilitate oriented immobilization on the surface of the sensor, in PBS was subsequently delivered to the chip for 17 minutes to facilitate passive adsorption of protein A to the native oxide on the surface of the silicon waveguide. In order to block any exposed sensor regions not covered by protein A, we delivered a 1 mg/mL solution of bovine serum albumin (BSA, Sigma-Aldrich A7030-10G, lot SLBD9162V) in PBS for 22 minutes. Next, a 20 µg/mL solution in PBS of a rabbit monoclonal antibody against the spike protein of the SARS-CoV-2 virus (Sino Biological 40150-R007, lot MA14MY2901) was flowed for 25 minutes to allow the antibodies to bind to the protein A immobilized on the waveguide. Finally, a 20 µg/mL solution of recombinant SARS-CoV-2 spike protein in PBS (Sino Biological 40591-V08H, lot LC16AP1501) was flowed for 17 minutes to facilitate spike protein analyte detection. PBS was flowed for 30 minutes at the end of the assay to again inspect the stability. Between each protein delivery step we flowed PBS for 10 minutes to rinse away any unbound protein and characterize resonance shifts introduced by the bound protein, independent of any differences in solution bulk refractive index. All flow steps were run at a flow rate of 30 μL/min except the antibody deposition, for which we used a flow rate of 20 μL/min. Like with the performance characterization, the fixed-wavelength readout was continually acquiring data at 1 Hz frequency throughout the assay.

The resulting fixed-wavelength readout data were linearized, and resonance peaks fitted and tracked as described above to extract the effective resonance peak shift as a function of time. The data were then corrected for sensor drift by least-squares fitting the initial 60-min PBS stability region of the equivalent peak shift data to a line, and subtracting that line from the full series of equivalent peak shift data to yield fit-corrected data. These fit-corrected data were subsequently analyzed using another custom semi-automated MATLAB® script that prompted the user to click on the PBS stability regions before and after each protein-delivery assay stage in order to calculate the average equivalent peak shift introduced by the protein binding. To calculate this shift, 150 fixed-wavelength equivalent peak shift datapoints were averaged in both PBS regions. These peak shift data were then tabulated and plotted in Figure 5.

## Acknowledgements

The authors are grateful to work and live on land that is the traditional, ancestral, and unceded territory of the Coast Salish Peoples, including the territories of the Musqueam, Squamish, and Tsleil-Waututh First Nations.

We would like to acknowledge Ahmad Sharkia for his work on the PDM design and Omid Esmaeeli for his assistance on the CMOS tapeout. We also thank So Jung (Elly) Kim and Yas Oloumi for their technical support in microfluidics and assay development. We are grateful to all members of our UBC silicon photonic biosensors team and Dream Photonics, Inc. for useful feedback and discussions. We also gratefully acknowledge helpful discussions with Jennifer Morales and Justin Bickford from the US Army Research Laboratory.

We thank the machine shop at the Stewart Blusson Quantum Matter Institute for their assistance in machining parts for the thermal stack, as well as 4D Labs and the SBQMI Advanced Nanofabrication Facility, where oxide-open fabrication was performed. This research was also supported in part through the computational resources and services provided by Advanced Research Computing at the University of British Columbia.

This material is based on research sponsored by the Air Force Research Laboratory under AIM Photonics (agreement number FA8650-21-2-1000). The U.S. Government is authorized to reproduce and distribute reprints for Governmental purposes notwithstanding any copyright notation thereon. The views and conclusions contained herein are those of the authors and should not be interpreted as necessarily representing the official policies or endorsements, either expressed or implied, of the United States Air Force, the Air Force Research Laboratory or the U.S. Government.

## Funding

This work was supported by the Schmidt Science Polymaths Program, the Optica Foundation Challenge Program, the Natural Sciences and Engineering Research Council of Canada (NSERC), Mitacs (through Accelerate and Elevate grants in partnership with Dream Photonics, Inc.), the Silicon Electronics-Photonics Integrated Circuits Fabrication (SiEPICfab) Consortium, CMC Microsystems, and Canada's Digital Technology Supercluster.

## Author contributions

K.C.C., L.C., and S.S. conceived and led the project. M.A.A-Q and S.M.G contributed to the concept, led the experimental design, led and conducted all experiments, and performed all data analyses. L.C., S.S., S.M.G., K.C.C., and M.A.A-Q invented the segmented sensor architecture. L.C. designed the photonic integrated circuits. M.A.A-Q and

S.S. conceived and designed the CMOS driver circuits and readout system. M.A.A-Q performed the simulations and derivations. M.M. and S.J.C developed and carried out the oxide-open fabrication process. S.M.G., K.N., S.K., and A.R. developed the microfluidics fabrication and integration process and fabricated the microfluidic devices. K.N., S.K., and A.R also contributed to microfluidic testing of sensor performance metrics and biomolecule detection. M.A.A-Q, S.M.G, A.R., Y.L., and P.T. contributed to the development of data analytics software used in this work. M.A.A-Q and S.M.G wrote the manuscript, and all authors reviewed and revised the manuscript.

## Competing interests

L.C., S.S., S.M.G., K.C.C., and M.A.A-Q are co-inventors on a patent describing this technology (US Patent No. 11940386, Application No. 17903931). L.C. and S.S. are co-founders of Dream Photonics, Inc. and S.M.G. and M.M. are employees of Dream Photonics, Inc.

# Supplementary Information: An integrated evanescent-field biosensor in silicon

Mohammed A. Al-Qadasi[*,1,#], Samantha M. Grist[*,1-4,#], Matthew Mitchell[4,5], Karyn Newton[1,3], Stephen Kioussis[1,3], Sheri J. Chowdhury[1,5], Avineet Randhawa[1-3], Yifei Liu[1,3], Piramon Tisapramotkul[1,3], Karen C. Cheung[1-3], Lukas Chrostowski[1,4,5], Sudip Shekhar[1,4,#]

[*] Contributed equally

[1] Department of Electrical and Computer Engineering, The University of British Columbia

[2] School of Biomedical Engineering, The University of British Columbia

[3] Centre for Blood Research, The University of British Columbia

[4] Dream Photonics, Inc.

[5] Stewart Blusson Quantum Matter Institute, The University of British Columbia

[#] Corresponding authors: alqadasi@ece.ubc.ca, sgrist@ece.ubc.ca, sudip@ece.ubc.ca

## Introduction

In this supplementary information, we describe the simulation results for the subwavelength grating waveguides used for sensing, along with the fabrication limitations of such designs (Supplementary Note 1). We describe the theoretical modeling of the bulk sensitivity and transmission for the segmented ring resonator (Supplementary Notes 2 and 3). The derivations of intrinsic and loaded quality factors for a segmented architecture are highlighted in Supplementary Note 4. We further analyze and characterize the temporal thermal response of the thermo-optic phase shifter (TOPS) used for sweeping and the resulting spatial thermal distribution across the SiP chip when the TOPS is activated (Supplementary Note 5). Supplementary Note 6 compares the measurement setups used in TLS and FW experiments and illustrates the photonic chip mounting setup. In Supplementary Note 7, we describe the data analysis and spectrum linearization for the FW interrogation scheme.

## Supplementary Note 1: Sub-wavelength grating design

In evanescent field-based silicon-photonic integrated designs, sensing is accomplished by the interaction of the analytes with the mode's effective refractive index. To enhance sensor sensitivity, maximizing the overlap of the evanescent field with the measured medium is essential. This improves the waveguide's susceptibility, which describes how the effective refractive index changes with variations in the cladding refractive index. Effective waveguide designs, such as strip TM [1], slot TE polarization [2], bimodal waveguides [3], and photonic crystal-based waveguides [4], are known for their strong mode overlap with the cladding. However, optimizing the geometry of such waveguide designs to enhance their sensitivity and minimize waveguide loss may not be feasible to fabricate in commercial foundry runs, for which stringent design rules need to be respected. This includes limitations on modifying the waveguide thickness or the minimum feature sizes. Subwavelength gratings were introduced as a means to tailor the waveguide optical properties by adjusting its effective refractive index [5]. By designing SWGs with grating periods significantly smaller than the wavelength of light, the waveguide material can effectively behave as a homogeneous medium [5]. There are multiple ways to simulate these SWG waveguides. First, in the equivalent index method, the refractive index of the SWG is approximated as the weighted average of the refractive indices of the grating material and the cladding, and a 2D mode solver is used to solve for the effective index. While this approximation helps to quickly simulate an SWG, its validity is constrained to designs with a grating periodicity much smaller than the wavelength of the input light (far from the Bragg condition) [6], which may not be possible for all foundry processes. To ensure accurate simulations for our design, we run 3D FDTD simulations [5].

To develop robust SWG waveguides for our sensing region of the segmented ring resonator architecture that met the design requirements of the AIM Photonics Base Active PIC process, we simulated the sensing performance fishbone SWG waveguides. For the SWG waveguide simulation, we perform FDTD band structure analysis on a single unit cell.

Bloch boundary conditions were set in the periodic direction, and PML layers were used on the remaining boundaries to absorb outward propagating waves and avoid reflections. With X, Y, Z used to denote the propagation, tangential, and normal axes to the surface of the chip, respectively, the FDTD window spans one SWG period along the x-axis, 6.5 µm along the y-axis and 1.72 µm along the z-axis. Fine meshing of dx = dy = dz = 0.01 µm is used to improve accuracy, and it was found to be important for the feature sizes of interest (e.g., the SWG period and the fishbone) to be an integer multiple of the mesh size. The simulated SWG waveguide uses a period of 0.2 µm and a fishbone width of 0.1 µm, and we simulated a range of duty cycles.

In our design, we employ a SWG waveguide with an optimized susceptibility of 0.83 RIU/RIU at a duty cycle of 0.5, providing a simulated bulk refractive index sensitivity of 63.1 nm/RIU in the segmented configuration. A fishbone-assisted structure was implemented to reinforce the SWG waveguide against delamination during or post-fabrication and ensure continuity in the effective index while transitioning from strip to SWG waveguides. This improves the fabrication yield and reduces optical losses, respectively.

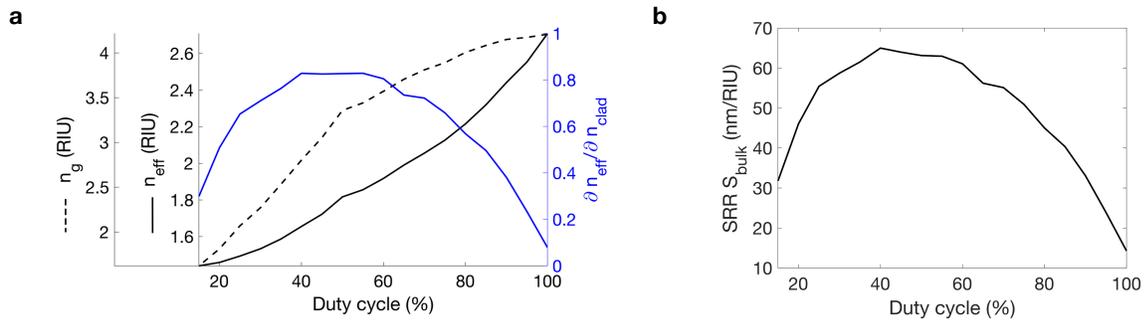

**Supplementary Figure 1. Sub-wavelength grating simulations used to inform SWG geometry selection.** (a) Simulated neff, ng, and waveguide susceptibility values for SWG values of different duty cycles, (b) Simulated bulk refractive index sensitivity for the segmented sensor design. The SWG incorporated in our design has a duty cycle of 0.5 which provides high susceptibility.

## Supplementary Note 2: Bulk refractive index sensitivity of segmented resonators

For a fixed wavelength sensor using a thermal phase shifter, the sensitivity can be represented in terms of the magnitude of shift in input electrical power, resulting phase shift, or equivalent resonance peak shift to the magnitude of change in the bulk refractive index of the waveguide cladding. In this section, we start by deriving the sensitivity of the SRR interrogated by a swept wavelength laser, $S_{bulk, TLS}[\frac{m}{RIU}]$. Next, we calculate the sensitivity given a fixed wavelength laser and a swept TOPS electrical power, $S_{bulk}[\frac{W}{RIU}]$. Finally, given the thermal efficiency of the heater, we calculate the equivalent wavelength shift and the corresponding bulk sensitivity which is equivalent to the derived TLS bulk sensitivity.

First, we derive the bulk sensitivity for a segmented resonator when a tunable input wavelength source is used. The bulk sensitivity of a resonator with multiple segments is proportional to the ratio of the weighted susceptibilities of the exposed waveguides to the weighted group indices of all segments. For a microring resonator, the sensitivity to changes in the cladding index is given in Supplementary Supplementary Equation (1) [7], where $\lambda_{res}$ represents the wavelength of resonance, $n_g$ represents the group index of the waveguide, and $\frac{\partial n_{eff}}{\partial n_{clad}}$ represents the susceptibility of the waveguide to changes in the refractive index of the cladding.

$$S_{bulk} = \frac{\Delta\lambda}{\Delta n_{clad}} = \frac{\lambda_{res}}{n_g}\frac{\partial n_{eff}}{\partial n_{clad}} \tag{1}$$

The resonance condition and the effective index of a waveguide as a function of an external perturbation, $\Delta n_{clad}$, are given in Supplementary Equation (2) and Supplementary Equation (3). The longitudinal mode number (order of interference) is represented by $m$. $L_T$ and $\beta$ represent the total length of the resonator and the fill factor (ratio of the exposed waveguide length to the overall length of the resonator), respectively.

$$\frac{m}{L_T} = \frac{n_{eff,e} \times \beta + n_{eff,u} \times (1-\beta)}{\lambda_{res}} \tag{2}$$

$$n_{eff} = n_g + \lambda \frac{\partial n_{eff}}{\partial \lambda} + \Delta n_{clad} \frac{\partial n_{eff}}{\partial n_{clad}} \tag{3}$$

The subscripts "$_e$" and "$_u$" in "$n_{eff,e}$", "$n_{eff,u}$", "$n_{g,e}$" and "$n_{g,u}$" are used to denote exposed and unexposed effective and group refractive indices. To derive the sensitivity, the modes before and after introducing the perturbation are equated as in Supplementary Equation (4). $\frac{\partial n_{eff}}{\partial n_{clad}}$ represents the waveguide's sensitivity (susceptibility) to the perturbation in the cladding, $\Delta n_{clad}$. The change in dispersion, $\frac{\partial n_{eff}}{\partial \lambda}$, is assumed negligible within the wavelength shift, $\Delta \lambda$.

$$\frac{[n_{g,e}+\lambda_{res}(\frac{\partial n_{eff}}{\partial \lambda})_e]\beta+[n_{g,u}+\lambda_{res}(\frac{\partial n_{eff}}{\partial \lambda})_u](1-\beta)}{\lambda_{res}} = \frac{[n_{g,e}+(\lambda_{res}+\Delta\lambda)(\frac{\partial n_{eff}}{\partial \lambda})_e + \Delta n_{clad}\frac{\partial n_{eff}}{\partial n_{clad}}]\beta+[n_{g,u}+(\lambda_{res}+\Delta\lambda)(\frac{\partial n_{eff}}{\partial \lambda})_u](1-\beta)}{(\lambda_{res}+\Delta\lambda)} \tag{4}$$

Simplifying Supplementary Equation (5), it follows that the bulk sensitivity ($S_{bulk}$) can be derived:

$$(\lambda_{res} + \Delta\lambda)[n_{g,e} + \lambda_{res}(\frac{\partial n_{eff}}{\partial \lambda})_e]\beta + (\lambda_{res} + \Delta\lambda)[n_{g,u} + \lambda_{res}(\frac{\partial n_{eff}}{\partial \lambda})_u](1-\beta) = \tag{5}$$

$$\lambda_{res}[n_{g,e} + (\lambda_{res} + \Delta\lambda)(\frac{\partial n_{eff}}{\partial \lambda})_e + \Delta n_{clad}\frac{\partial n_{eff}}{\partial n_{clad}}]\beta + \lambda_{res}[n_{g,u} + (\lambda_{res} + \Delta\lambda)(\frac{\partial n_{eff}}{\partial \lambda})_u](1-\beta)$$

$$\beta \cdot \Delta\lambda \cdot n_{g,e} + (1-\beta) \cdot \Delta\lambda \cdot n_{g,u} = \beta \Delta n_{clad} \cdot \lambda_{res} \cdot \frac{\partial n_{eff}}{\partial n_{clad}}$$

$$S_{bulk} = \frac{\Delta\lambda}{\Delta n_{clad}} = \frac{\beta \cdot \lambda_{res}}{\beta \cdot n_{g,e}+(1-\beta) n_{g,u}} \cdot \frac{\partial n_{eff}}{\partial n_{clad}} = \frac{\beta \cdot n_{g,e}}{\beta \cdot n_{g,e}+(1-\beta) n_{g,u}} \cdot \frac{\lambda_{res}}{n_{g,e}} \frac{\partial n_{eff}}{\partial n_{clad}}$$

For biosensor resonators with totally exposed waveguides, the sensitivity to perturbations in the cladding index, $\Delta n_{clad}$, can be shown as Supplementary Equation (6), with a sensitivity scaling factor $\rho = 1$. $n_{g,e}$ represents the group index of the exposed waveguide.

$$S_{bulk} = \frac{\Delta\lambda}{\Delta n_{clad}} = \rho \cdot \frac{\lambda_{res}}{n_{g,e}} \frac{\partial n_{eff}}{\partial n_{clad}} \tag{6}$$

For the SRR with the partially exposed resonators, $\rho$ represents the sensitivity scaling factor due to the partial exposure of the resonator waveguide and is no longer equal to 1. By extending Supplementary Equation (5) to account for multiple unexposed segments but assuming a single exposed segment, we can express $\rho$ as a function of the fill factor, $\beta$ (ratio of the exposed portion to the total length of the resonator) and the group indices ($n_g$) of exposed and unexposed waveguides. The factor $\rho$ represents the ratio of exposed to unexposed optical path length within the resonator (the optical path length fill factor of the resonator), and is given in Supplementary Equation (7). The factor $\eta_i$ represents the ratio of the unexposed waveguide segment $L_i$ to the overall length of the resonator, $L_T$, and $n_{g,i}$ represents the group index of the unexposed segment $L_i$

$$\rho = \frac{\beta \cdot n_{g,e}}{\beta \cdot n_{g,e} + \sum_{i=1}^{N} \eta_i n_{g,i}} \tag{7}$$

Intuitively, it can be inferred from Supplementary Equation (7) that the optical length fill factor can be simplified as the physical length fill factor, $\rho = \beta$, when there isn't a big variation in the group indices among the different segments of the resonator.

Using the same derivation method, we derive a generic expression of the bulk sensitivity for multiple ($M$) exposed waveguide segments as shown in Supplementary Equation (8). It can be observed that the bulk sensitivity is proportional to the ratio of the weighted susceptibilities of exposed waveguides to the weighted group indices of all $N$ (with $N > M$) resonator segments. $S_{wg,i,e}$ represents the susceptibility (i.e., $\frac{\partial n_{eff}}{\partial n_{clad}}$) of the exposed portion of waveguide "i" while $\eta_{i,e}$ represents the ratio of the length of each exposed waveguide to the overall length of the resonator. Therefore, the bulk sensitivity can be calculated and compared using Supplementary Equation (8).

$$S_{bulk} = \lambda_{res} \cdot \frac{\sum_{i=1}^{M} S_{wg,i,e} \eta_{i,e}}{\sum_{j=1}^{N} \eta_j n_{g,j}} \quad (8)$$

In the segmented sensor architecture using fixed-wavelength readout, we can visualize the readout approach as using the thermo-optic phase shifter (heater) to essentially compensate for the phase shift introduced by the change in cladding refractive index, bringing the resonance wavelength back to the fixed-wavelength readout wavelength. The heater power required to do this varies depending on the magnitude of the phase shift introduced by the binding event or bulk refractive index change. With this visualization in mind, we can derive the sensitivity in units of Watts/RIU. To do this, we derive the sensitivity by calculating the round-trip phase shift as in Supplementary Equation (9), where $\Phi_0$ represents the initial round-trip phase, $\Phi_{\Delta n_{clad}}$ represents the round-trip phase as a function of a change in the refractive index of the cladding, $\lambda_{res}$ represents the wavelength of resonance, $n_{eff}$ represents the effective refractive index, $n_g$ represents the group index, $L_e$ represents the length of the sensing segment, $\Delta n_{clad}$ represents the change in the refractive index of the cladding and $\Delta P_{heater}$ represents the heater electrical power. The indices $i$, $j$, and $k$ iterate over the $M$ exposed, $N$ unexposed, and $P$ total segments that make up the resonator, respectively, while the subscript $H$ denotes the heater segment.

$$\Phi_0 = \frac{2\pi}{\lambda_{res}} \left[ \sum_{i=1}^{M} n_{eff,i} L_{e,i} + \sum_{j=1}^{N} n_{eff,j} L_j + n_{eff,H} L_H \right]$$

$$\Phi_0 = \frac{2\pi}{\lambda_{res}} \left[ \sum_{i=1}^{M} (n_{g,i} + \lambda \frac{\partial n_{eff,i}}{\partial \lambda}) L_{e,i} + \sum_{j=1}^{N} (n_{g,j} + \lambda \frac{\partial n_{eff,j}}{\partial \lambda}) L_j + (n_{g,H} + \lambda \frac{\partial n_{eff,H}}{\partial \lambda}) L_H \right]$$

$$\Phi_{\Delta n_{clad}} = \frac{2\pi}{\lambda_{res}} \left[ \sum_{i=1}^{M} (n_{g,i} + \lambda \frac{\partial n_{eff,i}}{\partial \lambda} + \Delta n_{clad} \frac{\partial n_{eff,i}}{\partial n_{clad}}) L_{e,i} + \sum_{j=1}^{N} (n_{g,j} + \lambda \frac{\partial n_{eff,j}}{\partial \lambda}) L_j + (n_{g,H} + \lambda \frac{\partial n_{eff,H}}{\partial \lambda} - \Delta P_{heater} \frac{\partial n_{eff,H}}{\partial P_{heater}}) L_H \right]$$

Equating the phase shifts before and after the cladding change (assuming that the heater power is compensating for the cladding change to maintain constant $\lambda_{res}$):

$$\Phi_0 = \Phi_{\Delta n_{clad}}$$

$$0 = \Delta n_{clad} \sum_{i=1}^{M} \frac{\partial n_{eff,i}}{\partial n_{clad}} L_{e,i} - \Delta P_{heater} \frac{\partial n_{eff,H}}{\partial P_{heater}} L_H$$

$$S_{bulk}\left[\frac{W}{RIU}\right] = \frac{\Delta P_{heater}}{\Delta n_{clad}} = \frac{1}{L_H} \frac{1}{\frac{\partial n_{eff,H}}{\partial P_{heater}}} \sum_{i=1}^{M} \frac{\partial n_{eff,i}}{\partial n_{clad}} L_{e,i}$$

$$S_{bulk}\left[\frac{W}{RIU}\right] = \frac{L_T}{L_H} \frac{1}{\frac{\partial n_{eff,H}}{\partial P_{heater}}} \sum_{i=1}^{M} \left(\frac{\partial n_{eff,i}}{\partial n_{clad}}\right)_i \eta_{e,i} \quad (9)$$

The equivalent sensitivity for a tunable wavelength sensing can be calculated as:

$$S_{bulk}[\tfrac{m}{RIU}] = \tfrac{\Delta\lambda}{\Delta P_{heater}} \times S_{bulk}[\tfrac{W}{RIU}]$$

where $\tfrac{\Delta\lambda}{\Delta P_{heater}}$ represents the efficiency of the thermal phase shifter. With $\tfrac{\Delta\lambda}{\Delta P_{heater}} = \tfrac{FSR}{\Delta P_{2\pi}}$ and FSR given as:

$$FSR = \frac{\lambda_{res}^2}{\sum_{i=1}^{M} n_{g,i} L_i + \sum_{j=1}^{N} n_{g,j} L_j + n_{g,H} L_H}$$

By substituting this definition of phase shifter efficiency, we find $S_{bulk}[\tfrac{m}{RIU}]$:

$$S_{bulk}[\tfrac{m}{RIU}] = \frac{\lambda_{res}^2}{\sum_{i=1}^{M} n_{g,i} L_i + \sum_{j=1}^{N} n_{g,j} L_j + n_{g,H} L_H} \frac{1}{\Delta P_{heater}} \times S_{bulk}[\tfrac{W}{RIU}] \tag{10}$$

We can further simplify by multiplying Supplementary Equation (10) by the unity factor $\tfrac{\Delta n_{eff,H}}{\Delta n_{eff,H}}$, where $\Delta n_{eff,H}$ represents the change in the effective index induced by the thermal phase shifter to impose a 2π shift. Given that the phase shift difference before and after tuning the thermal phase shifter is 2π, it follows that:

$$\tfrac{2\pi}{\lambda_{res}}(n_{eff,H} + \Delta n_{eff,H}) L_H - \tfrac{2\pi}{\lambda_{res}}(n_{eff,H}) L_H = 2\pi$$

$$\Delta n_{eff,H} = \tfrac{\lambda_{res}}{L_H}; \quad \tfrac{\Delta n_{eff,H}}{\Delta n_{eff,H}} = \Delta n_{eff,H} \tfrac{L_H}{\lambda_{res}}$$

Then, Supplementary Equation (10) can be simplified:

$$S_{bulk}[\tfrac{m}{RIU}] = \frac{\lambda_{res}^2}{\sum_{i=1}^{M} n_{g,i} L_i + \sum_{j=1}^{N} n_{g,j} L_j + n_{g,H} L_H} \frac{1}{\Delta P_{heater}} \cdot S_{bulk}[\tfrac{W}{RIU}] \cdot \left(\tfrac{L_H}{\lambda_{res}} \cdot \Delta n_{eff,H}\right) \tag{11}$$

$$= \frac{\lambda_{res} L_H}{\sum_{i=1}^{M} n_{g,i} L_i + \sum_{j=1}^{N} n_{g,j} L_j + n_{g,H} L_H} \frac{\Delta n_{eff,H}}{\Delta P_{heater}} \cdot S_{bulk}[\tfrac{W}{RIU}]$$

$$= \frac{\lambda_{res} L_H}{\sum_{i=1}^{M} n_{g,i} L_i + \sum_{j=1}^{N} n_{g,j} L_j + n_{g,H} L_H} \frac{\Delta n_{eff,H}}{\Delta P_{heater}} \cdot \left[\tfrac{L_T}{L_H} \tfrac{1}{\tfrac{\partial n_{eff,H}}{\partial P_{heater}}} \sum_{i=1}^{M} S_{wg,i} \eta_{s,i}\right]$$

Approximating $\tfrac{\partial n_{eff,H}}{\partial P_{heater}} \approx \tfrac{\Delta n_{eff,H}}{\Delta P_{heater}}$:

$$S_{bulk}[\tfrac{m}{RIU}] = \frac{\lambda_{res} L_T}{\sum_{i=1}^{M} n_{g,i} L_i + \sum_{j=1}^{N} n_{g,j} L_j + n_{g,H} L_H} \left[\sum_{i=1}^{M} S_{wg,i} \eta_{s,i}\right] \tag{11a}$$

Supplementary Equation 11a can be further simplified into a similar format as Supplementary Equation 8 by considering the individual waveguide fill factors of each of the *P* total (both exposed and unexposed) segments of the resonator:

$$S_{bulk}[\tfrac{m}{RIU}] = \frac{\lambda_{res}}{\sum_{k=1}^{P} n_{g,k} \eta_k} \left[\sum_{i=1}^{M} S_{wg,i} \eta_{s,i}\right] \tag{11b}$$

Therefore, knowing the group indices, susceptibility, and fill factors of the various waveguide segments enables the calculation of the bulk sensitivity of an SRR implementation for either interrogation scheme.

## Supplementary Note 3: Segmented ring resonator (SRR) transmission

To validate and understand the proposed design, we derive the transmission equation for the segmented resonator architecture described in Fig. 2. This enables us to compare simulated results with measurements and accurately extract losses from measurement data. For a ring resonator incorporating an interferometric coupler, the matrix transfer function can be calculated as in Supplementary Equation (12), where $e^{-\beta}$ represents the average loss within the MZI arms and $e^{\pm\Delta\beta}$ stands for the loss mismatch between both arms. Assuming equal field coupling, $k$, and transmission, $r$, ratios such that $r = k = \frac{1}{\sqrt{2}}$, the intensity transfer function of the asymmetric MZI coupler can be simplified as in Supplementary Equation (14). $T^H$ denotes the Hermitian (i.e., complex conjugate transpose) of matrix $T$. The subscripts "$_E$" and "$_I$" in $T_E$ and $T_I$ denote the transmission of the electric field and intensity, respectively.

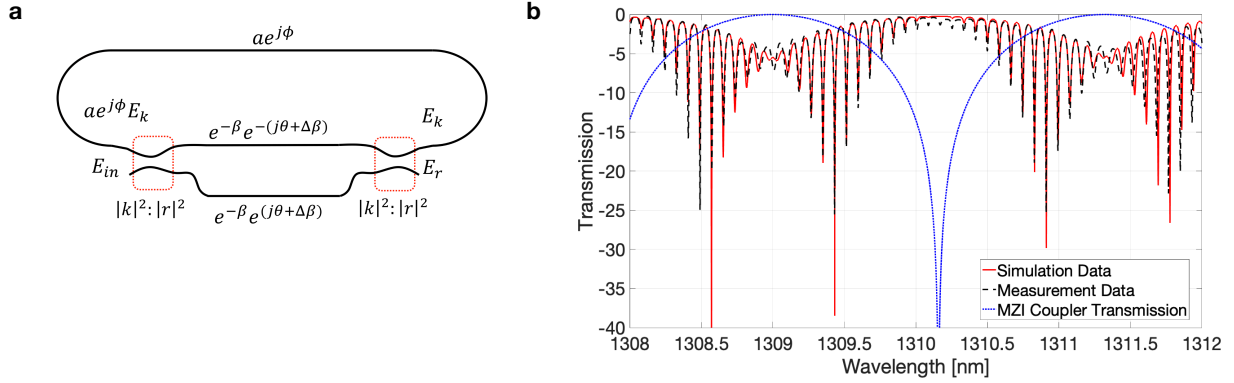

**Supplementary Figure 2. Experimentally validated modeling of segmented ring resonator transmission.** (a) A simplified illustration of a resonator with interferometric coupling, illustrating model parameters of interest. (b) Modeled (Supplementary Equation 15) and measured transmission of a segmented ring resonator, along with simulated transmission for the incorporated MZI coupler (Supplementary Equation 13). The fitting parameters used for the modeled transmission, with $R^2 = .70547$, are as follows: a = 0.5466 m$^{-1}$, $k = r = 0.7071$, $n_{eff,MZI} = 2.7$, $\Delta L_{MZI} = 136.96$ μm, $n_{eff,SRR} = 3.9$ and $L_{SRR} = 5.3$ mm. The resonance round-trip phase shift is given as $\Phi = \frac{2\pi n_{eff,MZI} \Delta L_{MZI}}{\lambda}$, while the phase shift due to the MZI coupler is given as $\theta = \frac{2\pi n_{eff,SRR} L_{SRR}}{\lambda}$.

$$T_{E,MZI} = e^{-\beta}\begin{bmatrix} r & ik \\ -ik & r \end{bmatrix}\begin{bmatrix} e^{j\theta+\Delta\beta} & 0 \\ 0 & e^{-(j\theta+\Delta\beta)} \end{bmatrix}\begin{bmatrix} r & ik \\ -ik & r \end{bmatrix} = e^{-\beta}\begin{bmatrix} r^2 e^{j\theta+\Delta\beta}+k^2 e^{-(j\theta+\Delta\beta)} & ikr(e^{j\theta+\Delta\beta}+e^{-(j\theta+\Delta\beta)}) \\ -ikr(e^{j\theta+\Delta\beta}+e^{-(j\theta+\Delta\beta)}) & k^2 e^{j\theta+\Delta\beta}+r^2 e^{-(j\theta+\Delta\beta)} \end{bmatrix} = \begin{bmatrix} T_{E,11} & T_{E,12} \\ T_{E,21} & T_{E,22} \end{bmatrix} \quad (12)$$

$$T_{I,MZI} = T_{E,MZI}(T_{E,MZI})^H = \frac{e^{-2\beta}}{2}\begin{bmatrix} \cosh(2\Delta\beta)+\cos(2\theta) & i[\cosh(2\Delta\beta)+\cos(2\theta)] \\ -i[\cosh(2\Delta\beta)+\cos(2\theta)] & \cosh(2\Delta\beta)+\cos(2\theta) \end{bmatrix} \quad (13)$$

Supplementary Equation (12) and Supplementary Equation (13) illustrate that adjusting the coupling ratio of the input and output directional couplers, along with managing phase and loss mismatches in the MZI arms, effectively controls the output swing of an MZI. This swing defines the coupling range of the MZI-coupled segmented resonator. By optimizing these parameters to account for variations in resonator loss and fabrication, the transmission range of the segmented resonator can be adjusted to extend around critical coupling.

The field transfer function of an SRR with an MZI coupler is given in Supplementary Equation (14). With negligible loss mismatch in the MZI coupler, the intensity transfer function for an asymmetric MZI simplifies to Supplementary Equation (15), where $a$ represents the field transmission in the resonator. To validate our proposed design model, Supplementary Fig. 2 presents the measured transmission spectra using $P_{laser} = 0\ dBm$, along with the transmission model (Supplementary Equation (15)) , showing good agreement.

$$T_{E,SRR} = \frac{E_r}{E_{in}} = \frac{T_{E,22}-ae^{j\Phi}|T_E|}{1-ae^{j\Phi}T_{E,11}} \quad (14)$$

$$T_{I,SRR} = T_{E,SRR}T_{E,SRR}^* \approx \frac{a^2[k^4+r^4+2k^2r^2\cos(2\theta)+4k^2r^2\sin^2(\theta)]^2+2k^2r^2\sin^2(\theta)+4a[k^4+r^4+2k^2r^2\cos(2\theta)+4k^2r^2\sin^2(\theta)][kr\sin(\theta)]\cos(\Phi)}{1+4k^2r^2\sin^2(\theta)+4akr\sin(\theta)\cos(\Phi)} \quad (15)$$

## Supplementary Note 4: Quality factor of segmented resonators

The quality factor describes the sharpness of resonance relative to the central wavelength and is inversely proportional to the round-trip losses of a resonator. Calculating the quality factor provides insight into the contribution of segment losses on the limit of detection, highlighting the advantage of employing low-loss multimode waveguides for efficient routing. For an SRR design, the loaded quality factor, $Q_{loaded}$, is defined as follows [8]:

$$Q_{loaded} = \left(\frac{1}{Q_{int}} + \frac{1}{Q_c}\right)^{-1} \tag{16}$$

where $Q_{int}$ and $Q_c$ represent the quality factors due to the intrinsic and coupling losses of the resonator, respectively. $N$, $T_R$ and $L_R$ are used to represent the number of segments, total round-trip time and length of the ring resonator, respectively. With $\alpha_i$, $L_i$ and $t_i$ denoting the rate of loss, length, and propagation time of the $i^{th}$ segment, Supplementary Equation (17) derives the intrinsic quality factor for a segmented resonator from that for a classical resonator [8]. $\omega_0$ denotes the angular frequency of the resonance and $E_{SRR}[n]$ represents the electric field circulating within the resonator in the nth cycle [9].

$$|E_{SRR}[n]|^2 = |E_{SRR}[n-1]|^2 e^{-\sum_{i=1}^{N} \alpha_i L_i} \tag{17}$$

$$d|E_{SRR}[n]|^2/dn = -|E_{SRR}[n]|^2 \sum_{i=1}^{N} \alpha_i L_i$$

$$d|E_{SRR}[n]|^2/dt = (1/T_R)|E_{SRR}[n]|^2 dn$$

$$d|E_{SRR}[n]|^2/dt = -(|E_{SRR}[n]|^2 \sum_{i=1}^{N} \alpha_i L_i)/T_R$$

$$Q_{int} = \omega_0 \frac{|E_{SRR}|^2}{-\partial|E_{SRR}|^2/\partial t} = \omega_0 \frac{|E_{SRR}|^2 \times T_R}{|E_{SRR}|^2 \sum_{i=1}^{N} \alpha_i L_i} = \omega_0 \frac{\sum_{i=1}^{N} t_i}{\sum_{i=1}^{N} \alpha_i L_i}$$

The propagation time through each segment can be presented in terms of its length, $L_i$, and group velocity, $v_{g,i} = c/n_{g,i}$. This simplifies the expression as a function of the fill factors $\eta_i$ of each segment as shown into Supplementary Equation (18).

$$Q_{int} = \omega_0 \frac{\sum_{i=1}^{N} \frac{n_{g,i} L_i}{c}}{\sum_{i=1}^{N} \alpha_i L_i} = \frac{2\pi}{\lambda} \frac{\sum_{i=1}^{N} n_{g,i} \eta_i}{\sum_{i=1}^{N} \alpha_i \eta_i} \tag{18}$$

Similarly, the quality factor due to coupling is calculated in Supplementary Equation (19).

$$Q_c = \omega_0 \frac{|E_{SRR}|^2}{-2\ln(t) \times |E_{SRR}|^2/T_R} = \frac{2\pi}{\lambda} \frac{\sum_{i=1}^{N} n_{g,i} \eta_i}{-2\ln(|r|)/L_R} \tag{19}$$

Where $|r|$ represents the field transmission of the coupler. For MZI coupling, this is equal to the MZI transmission, given in Supplementary Equation (14), such that $r = \sqrt{T_{11,MZI}} = \sqrt{T_{22,MZI}}$. Therefore, the loaded quality factor is given in Supplementary Equation (20).

$$Q_{loaded} = \frac{2\pi}{\lambda} \times \frac{\sum_{i=1}^{N} n_{g,i} \eta_i}{\sum_{i=1}^{N} \alpha_i \eta_i - 2\ln(|r|)/L_R} \tag{20}$$

The critically coupled quality factor, where coupling is matched with resonator losses, is presented in Supplementary Equation (21).

$$Q_{critical} = \frac{\pi}{\lambda} \times \frac{\sum_{i=1}^{N} n_{g,i} \eta_i}{\sum_{i=1}^{N} \alpha_i \eta_i} \quad (21)$$

## Supplementary Note 5: Thermal characterization of TOPS

The high thermo-optic coefficient of silicon allows for temperature-based modulation of its refractive index. Resistive heaters are often used as thermo-optic phase shifters; however, the efficacy of modulation in terms of sweep range and speed depends significantly on the thermodynamic behavior of the system. In other words, thermal dissipation dictates the ON and OFF switching of these phase shifters. With slow and finite thermal dissipation, mounted silicon chips behave as low pass filters, attenuating electrical signals at high sweep rates and limiting the resonance sweep range [10].

We model the thermal dissipation by a Foster RC-network to investigate the transient response of the phase shifters for our mounted biosensor chip, where R and C denote the thermal resistance and capacitance values across the various stack layers (Supplementary Fig. 3a). This helps understand the tunable range that can be achieved for a sweep frequency for a given heater voltage range. In our measurements, we sweep the TOPS at various frequencies to capture the transient thermal response due to multiple thermal time constants. We use the measured heater resistance to calculate the input electrical power for a periodic voltage ramp signal which is then convolved with the transfer function of the thermal network and obtain the RC values that better match for the modeled transient thermal response with the measured data across all sweep frequencies (Supplementary Fig. 2c). These RC values are helpful later to calibrate for the equivalent wavelength axis, given that the TOPS bandwidth limits the sweep range.

In our modeling as well as thermal characterization, we use a voltage sweep range of 0 – 2.5 V, equivalent to that used by the CMOS drivers. With $P_{PS}$ representing the electrical power applied across the phase shifter, $\Phi_{PS}$ representing the corresponding optical phase shift, $\frac{\partial \Phi_{PS}}{\partial P_{PS}}$ representing the thermal efficiency of the TOPS derived from characterizing the resonator spectral response across various electrical power values and $\frac{\Delta P_{PS,0Hz}}{\Delta T_{PS,0Hz}}$ representing the steady-state temperature difference induced by the heater, we calculate the equivalent resonator sweep range shown in Supplementary Fig. 3c using Supplementary Equation (22). With a sweep frequency of 1 Hz, we obtain a sweep range of $8\pi$ which is used in our biosensor measurements (Supplementary Fig. 3c). $\Delta T(f)$ represents the temperature swing at sweep frequency $f$.

$$\Delta \Phi_{PS}(f) = \Delta T(f) \times \left(\frac{\Delta P_{PS,0Hz}}{\Delta T_{PS,0Hz}}\right) \cdot \left(\frac{\partial \Phi_{PS}}{\partial P_{PS}}\right) \quad (22)$$

An infrared camera (Jenoptik Variocam) was used for thermal imaging of the photonic chip, in order to characterize the heat distribution as a function of distance from the phase shifter and sweep frequency. Sweep frequencies of 0.1, 1, 10, 25, 50, and 100 Hz were used. Given that the capture rate of the thermal camera is limited to 25 frames per second (fps), we use temporal subsampling (i.e., undersampling or bandpass sampling [11]) to capture data at higher sweep frequencies.

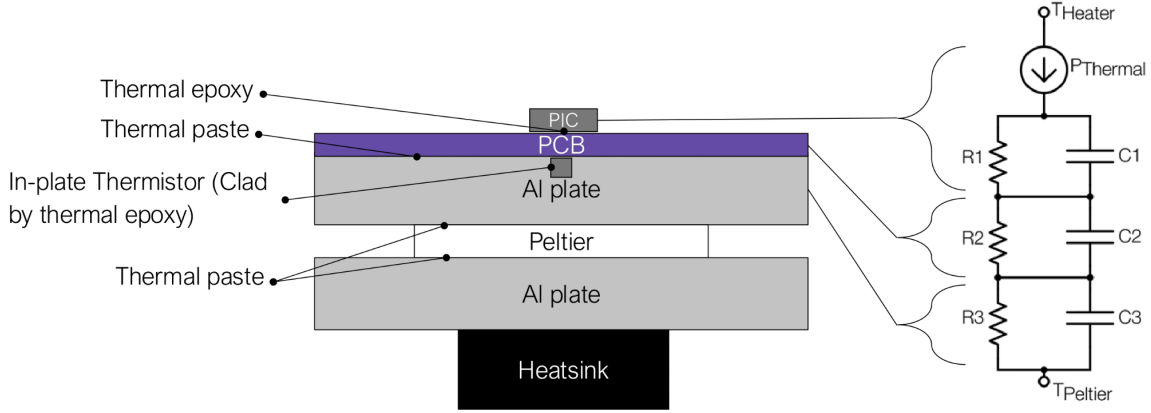

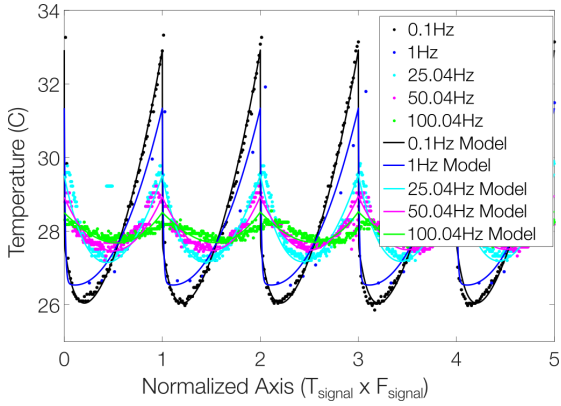
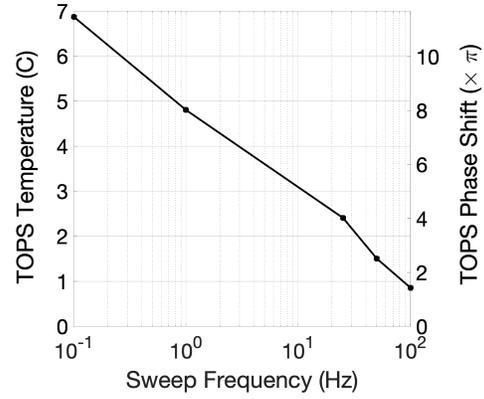

**Supplementary Figure 3. Experimental setup and results for frequency-dependent thermal characterization of thermo-optic phase shifter (TOPS) performance.** (a) A side-view illustration for the chip mount stack and the corresponding RC thermal dissipation network, (b) measured temperature values for the thermal phase shifter at different frequencies along with the simulated results for the fitted model for $R_1$= 200 K/W, $R_2$= 0.1 K/W, $R_3$= 1 K/W and $C_1$=50 µJ/K, $C_2$= 10 J/K, $C_3$= 1 J/K, and $\alpha$ = 0.1 K/W (Supplementary Equation (24)). The thermal dissipation model we have implemented aligns with the observed thermal behavior, which substantiates our transient response analysis for TOPS interrogated with a periodic signal. (c) Simulated TOPS peak-to-peak temperature swing and corresponding phase shifts for different sweep frequencies. In our work, we opted for 1 Hz, which ensures the capture of sufficient peaks per sweep for proper characterization and peak tracking (Supplementary Note 7). Given the relatively slow molecular binding and bulk refractive index shifts that we observe (e.g., in the main text Figures 4-5), this interrogation rate is adequately fast to enable acquiring details of the transient response as well as potentially perform signal averaging in the future, and exceeds the frequency of tunable laser source interrogation using our system by >10×.

In order to extract the thermal efficiency, $P_\pi$, of the heaters, we measure the steady state resonance shift of the resonator for various heater voltages. By fitting the resonance shift to a quadratic relation, $\Delta\lambda_{res} = \gamma V_{TOPS}^2$, we calculate the thermal efficiency of TOPS as shown in Supplementary Equation (23). $\gamma$ is a coefficient obtained from the characterized quadratic fit (Fig. 3a-iii) to be used along with the measured FSR and TOPS electrical resistance to obtain thermal efficiency of the TOPS from measurement data.

$$\frac{2P_\pi}{FSR} = \frac{P_{TOPS}}{\Delta\lambda_{res}} \tag{23}$$

$$P_{TOPS} = \frac{V_{TOPS}^2}{R_{TOPS}}$$

$$\rightarrow P_\pi = \frac{1}{2} \frac{FSR}{\Delta\lambda_{res}} P_{TOPS} = \frac{1}{2} \frac{FSR}{\Delta\lambda_{res}} \frac{V^2_{TOPS}}{R_{TOPS}} = \frac{1}{2} \frac{FSR}{\gamma R_{TOPS}}$$

With an average measured TOPS resistance of $R_{TOPS} \sim 40\Omega$ and fitted $\gamma \approx 1/20\ (nm/V^2)$ in our TOPS characterization (Fig. 3a-iii), we obtain $P_\pi = 21.25\ mW/\pi$. The transient temperature induced by the phase shifters is calculated in Supplementary Equation (24).

$$T_{TOPS,model} = T_{TOPS,min} + [\alpha(T_{TOPS,max} - T_{TOPS,min}) \times P_{TOPS}(t) * h_{TOPS}(t)]/[\int_0^\infty h_{TOPS}(t)dt] \qquad (24)$$

Here, the α factor describes the ratio of induced temperature change due to supplied electrical power to the TOPS, $\alpha = \Delta T/P_{TOPS}$, and is used to scale the input electrical power signal to the measured temperature readings. $h_{TOPS}(t)$ represents the impulse response transfer function of the TOPS and is given in Supplementary Equation (25). This is derived from thermal power dissipation models utilizing an RC circuit analogy [12,13].

$$h_{TOPS}(t) = \mathcal{L}^{-1}\left\{\frac{1}{\sum_{i=1}^{N} R_i} \sum_{i=1}^{N} \frac{R_i}{1+sR_iC_i}\right\} = \frac{1}{\sum_{i=1}^{N} R_i} \sum_{i=1}^{N} R_i e^{-\frac{t}{R_iC_i}} \qquad (25)$$

While thermo-optic phase shifters are widely used due to their maturity, ease of design, and availability in most MPW fabrication runs, several enhancements can be implemented to improve efficiency and reduce thermal crosstalk. These include integrating thermal insulation trenches [14,15]. Recent advancements in modulation technologies, such as polymer-based and liquid-crystal modulators, have also been demonstrated in SiP platforms. Adopting these technologies could significantly mitigate thermal crosstalk and reduce the required separation distance between the tuning and coupling regions of the SRR, while also enabling resonance interrogation at higher frequencies.

Supplementary Fig. 4 illustrates the circuit implementation used to convert a 10b digital ramp code into a 1b pulse-density modulated (PDM) signal. Driving a thermal phase shifter with this signal induces a phase shift equivalent to that induced by the analog value of the input digital code. The range of resonance shifts depends on the amplitude of the driving ramp signal and the thermal efficiency of the phase shifter. As the output of the PDM modulator swings between 0 V and the nominal supply voltage (i.e., 1 V), the resulting analog voltage amplitude is limited to 1 V. Therefore, the digital PDM signal is scaled to higher voltages using the circuit shown in Supplementary Fig. 4b. The signal pulse annotations reflect the signal scaling performed in this work, wherein the output of the CMOS driver is designed to output PDM pulses swinging from 0 - 5 V.

Three regions (0 - $V_{DD}$, 0 - $V_{DDH}$, and $V_{DDH}$ - $2V_{DDH}$) are highlighted in Supplementary Fig. 4b to illustrate the signal swing at different stages of the circuit. The first region [0 - $V_{DD}$] highlights the circuit components driven by digital signals within the nominal voltage supply, i.e., 0 - 1 V. The second region [0 - $V_{DDH}$] represents where the PDM signal is translated to a higher supply voltage denoted by $V_{DDH}$. Scaling the signal beyond this voltage requires level shifting and careful circuit and layout design. '$2V_{DDH}$' represents twice the $V_{DDH}$ supply voltage. Transistors with thick oxide layers (all transistors) and guard rings in the 0 - $V_{DDH}$ and $V_{DDH}$ - $2V_{DDH}$ regions were used to tolerate potential differences up to 3.3 V across each transistor.

The digital 1b signal output from the PDM modulator is first converted to a differential digital signal using a series of inverters then fed to the input NMOS transistors of the level translator, [M1, M1']. When the PDM signal switches from 0 - 1 V, the NMOS transistors M1 and M1' turn ON and OFF, respectively. The cross-coupled PMOS transistors [M3, M3'] represent active pull-up loads with regenerative gain to speed up the switching. With M1 turned ON, the voltage across the node V1 starts to drop. When V1 reaches $V_{DDH} - |V_{THP}|$, where $|V_{THP}|$ represents the threshold voltage of the PMOS transistors, it causes M3' to switch ON and begin charging up the node V1'. The regeneration, which describes the phase at which the cross coupled transistors start amplifying the differential voltage across their drains, is attained when the nodes V1 and V1' have equal voltages. With M1 and M3 switched ON at the same time prior to regeneration, the speed of the transition is dictated by the pulldown of V1 and the charge–up of node V1'. Therefore, the input NMOS M1 and M1' are designed to be much larger than the regenerative pair transistors. Furthermore, cross-coupled NMOS drain-source connected transistors are added to introduce dynamic current which helps speed up the charging and discharging of the nodes across the drains of the PMOS transistors. This assists the voltage translation across the slow-fast and fast-slow PVT corners where the NMOS and PMOS transistors are weak enough to pull down or charge up the voltages across their nodes causing slow transitions. Once regeneration starts, the voltage V1 drops to 0 V while voltage V1' approaches $V_{DDH} - |V_{THP}|$ and higher causing M3 to switch OFF. Inverters [I1, I1'] and [I2, I2'] are used to buffer the 0 - $V_{DDH}$ PDM signal.

The third region [$V_{DDH}$ - $2V_{DDH}$] comprises back to back inverters I4 and I4' which hold the signal between voltages $V_{DDH}$ and $2V_{DDH}$. The voltages in nodes V3 and V3' represent the level-shifted versions of nodes V2 and V2' and should toggle simultaneously. This is achieved using the common-gate-biased PMOS and NMOS transistors in I3 and I3', in addition to the decoupling capacitors C1 and C2. When the voltage V2 transitions from $V_{DDH}$ to 0 V (e.g., from 2.5 to 0 V), V3 transitions from $2V_{DDH}$ to $V_{DDH}$ (e.g. from 5 to 2.5V). At V2 = $V_{DDH}$ and V2' = 0 V, the NMOS of I3 and PMOS of I3' are OFF, while the PMOS of I3 and NMOS of I3' are ON. As V2 drops below $V_{DDH} - V_{THN}$, where $V_{THN}$ represents the threshold voltage of the NMOS transistors, the NMOS of I3 turns ON creating a path for node V3 to discharge. On the other hand, the NMOS transistors of I3' switches off only when the voltage in V2' reaches $V_{DDH} - V_{THN}$. Therefore, the decoupling capacitors C1 and C2 are incorporated to create paths for the V3 and V3' to charge and discharge and ensure simultaneous transition of the low and high shifted signals. When the nodes V3 and V3' reach the same voltage, the back-to-back inverters start the regeneration. The stack of level shifter (0 - $V_{DDH}$), common-gate-based I3 transistors and back-to-back I4 inverters are designed to achieve truly differential cascode drivers[16]. Since the thermal phase shifter has low resistance ~40 Ω, the front-end driving transistors, M7 - M12 (Figure 4c), are scaled up to achieve the desired current and ON resistances. The chain of inverters [I5, I5'] and [I6, I6'] are used for tapered buffering.

A 3-tap termination with series P+ polysilicon resistors was incorporated in the front-end drivers as shown in Supplementary Fig. 3b, where $R_{n0} = R_{p0}$ = 24.6 Ω , $R_{n1} = R_{p1}$ = 110.6 Ω, $R_{n2} = R_{p2}$ = 295 Ω. Switching transistors, M7 - M12, were sized to provide total series resistance of 30 Ω. This yields series resistance in the range of ~ 35 - 325 Ω. This tunability is added to be able to drive TOPS with different resistance values in which the voltage swing across TOPS is controlled by the divided voltage between the series resistance values. The swing of the switching signals SNx and SPx is scaled using similar level translation and level shifting circuits to match the voltage ranges of the switch transistors. Supplementary Fig. 3c illustrates the cross-sectional view of the front-end transistors. Multiple layers of guard rings were incorporated to enhance the isolation of the high-swing PDM signal from the substrate and to avoid forward biasing the p-substrate n-well diode, which might trigger latch-up. These guard rings also play a crucial role in mitigating breakdown in the doped regions by reducing the large potential difference across the guard rings.

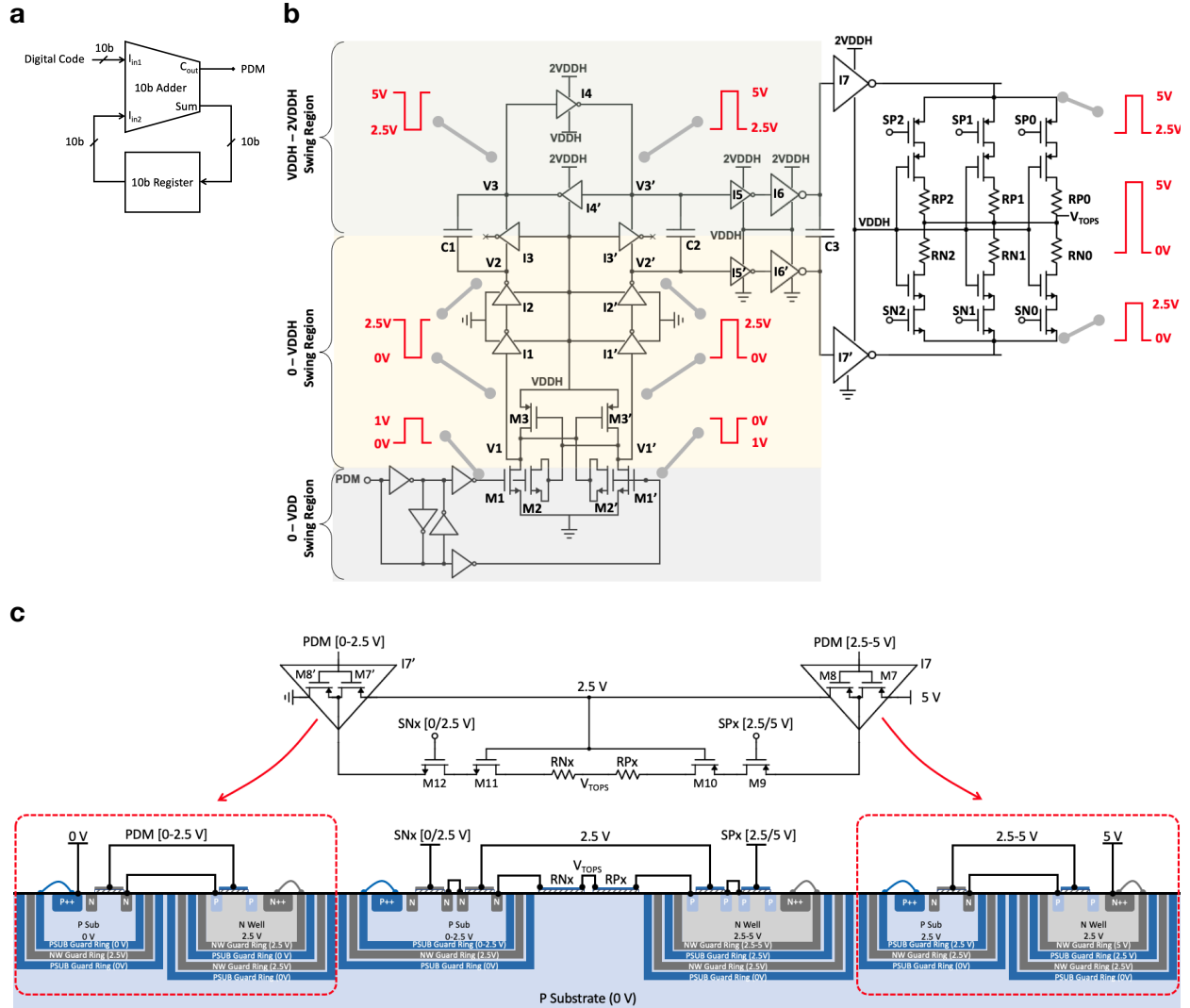

**Supplementary Figure 4. Illustration of the CMOS driver circuits.** (a) Schematic of the PDM modulator, (b) Schematic of the voltage level translator with the PDM signal at different nodes of the circuit, (c) Cross-sectional layout view of the front-end drivers for $V_{DDH}$ = 2.5 V.

## Supplementary Note 6: Fixed-wavelength measurement setup

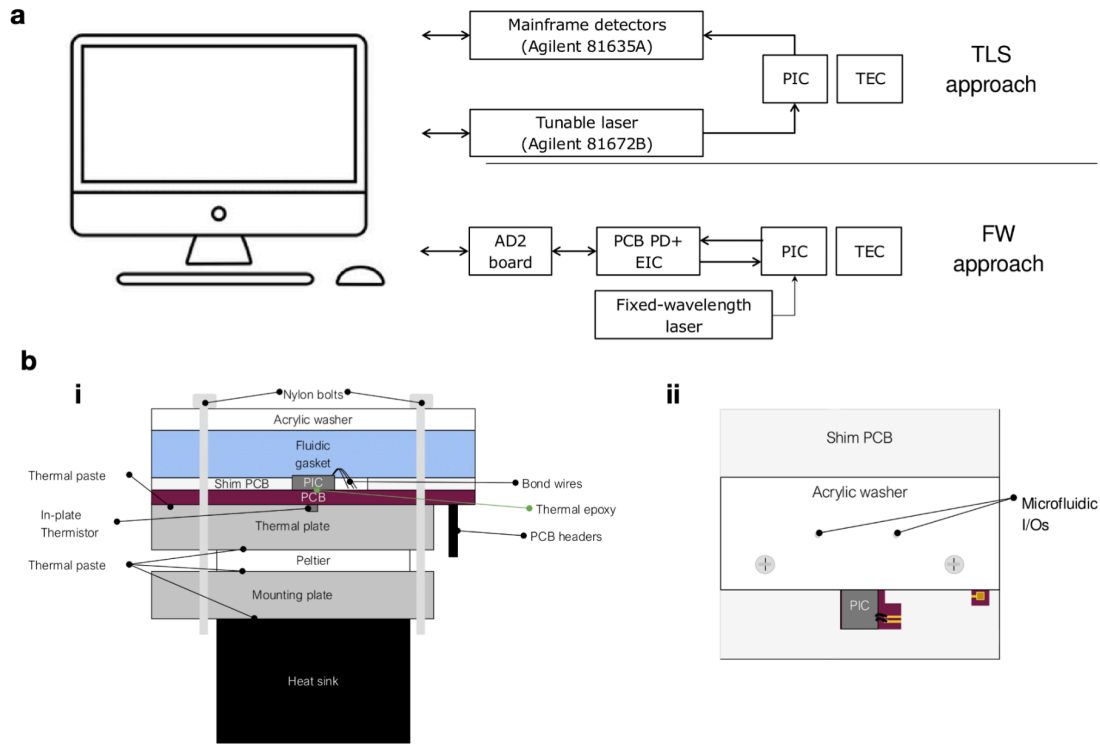

**Supplementary Figure 5. Measurement setup details for SRR characterization experiments.** (a) The proposed resonance interrogation scheme uses a fixed-wavelength source, compared to the conventional tunable laser source setup. For the proof of concept demonstration in this work, both setups incorporate an SRC LDC for thermal stabilization of the photonic chip and a PC for data logging. In the FW implementation, the EIC settings are configured using the AD2 board. (b) The PIC mounting setup, depicting the thermal control heat sink, Peltier, and feedback thermistor, as well as the carrier PCB, mounting shim, PIC, wire bonds, microfluidic gasket, and acrylic washer. (i) depicts the cross-sectional view while (ii) depicts the top view.

## Supplementary Note 7: Data analysis and spectrum linearization details

Since the relationship between the input signal and the resonance sweep is nonlinear as a function of input voltage and frequency-dependent, we linearize our fixed-wavelength sensor's transmitted intensity vs. time/voltage data in two steps: (1) input regulation and (2) a linear time-invariant (LTI) transfer function. The input regulation step describes the quadratic behavior of the resistive TOPS. The input voltage, $V_{input}$, generates thermal heat which is linear with the dissipated electrical power ($V^2_{input}/R_{TOPS}$). Using the input voltage signal and the measured value of the TOPS resistance, the electrical power is calculated. The characterized heater efficiency is used to scale the applied electrical power to the corresponding equivalent resonance shift as in Supplementary Equation (26), which is then convolved with a thermal dissipation transfer function given in Supplementary Equation (27). This transfer function uses an array of R and C values describing the thermal resistivities and capacitances of the different layers between TOPS and the Peltier module. Since the thermal efficiency has already been taken into account in Supplementary Equation (26), the thermal dissipation LTI function, h(t), should not introduce any further scaling for the thermal efficiency. Since this transfer function is normalized by dividing by the area of the transfer function, the R and C values don't necessarily need to represent the actual R and C values, but any scaled version of them can be used. We start by using arbitrary R and C values such as the ones used to model the temperature, then tune the R and C values until the spectrum peaks are equally spaced and have the same FSR as that measured with TLS readout.

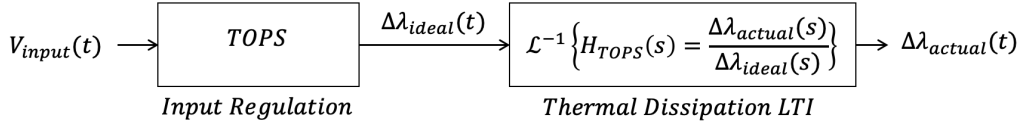

**Supplementary Figure 6 Illustration of the data linearization workflow, incorporating the characteristics of the TOPS and thermal dissipation behavior.**

$$\Delta\lambda_{ideal}(t) = \frac{FSR}{P_{2\pi}} \frac{V^2_{input}(t)}{R_{TOPS}} \tag{26}$$

$$H_{TOPS}(s) = \frac{\Delta\lambda_{Thermal-RC}(s)}{\Delta\lambda_{ideal}(s)} = \frac{1}{\sum_{i=1}^{N} R_i} \sum_{i=1}^{N} \frac{R_i}{1+sR_iC_i} \tag{27}$$

$$\Delta\lambda_{actual}(t) = \Delta\lambda_{ideal}(t) * h_{TOPS}(t)/[\int_0^\infty h_{TOPS}(t)dt] \tag{28}$$

Supplementary Fig. 7 depicts the analysis workflow used to extract the equivalent resonance peak shift vs. time data. Initially, the acquired optical power vs. time data are sliced into individual pre-spectra. Using the data linearization illustrated in Supplementary Fig. 6, the sliced spectra are linearized into optical power vs. equivalent wavelength shift spectra. This is done by expanding the poles in Supplementary Equation (27), obtaining the numerator and denominator of the LTI transfer function and calculating the calibrated resonance shift axis, $\Delta\lambda_{actual}$. All resonance peaks of the spectrum are fitted to a Lorentzian function. The peak position, full-width at half maximum (FWHM), quality factor and FSR is calculated for each peak.

With the data linearized, we use a modified Lorentzian model, Supplementary Equation (29), to fit the peaks where $\lambda_0$ represents the wavelength of resonance, $T_0$ represents the baseline intensity, $\gamma$ represents the slope of the baseline (contributed by both the the interferometric coupling as well as the grating coupler transmission profile), and $\beta$ represents the peak height.

$$f(\lambda_0, T_0, \gamma, \beta, FWHM) = T_0 + \gamma(\lambda - \lambda_0) + \frac{\beta(FWHM/2)^2}{(\lambda - \lambda_0)^2 + (FWHM/2)^2} \tag{29}$$

We extract the vector of 5 parameters, $\hat{v} = [\lambda_0, T_0, \gamma, \beta, FWHM]$, of the Lorentzian function described in Supplementary Equation (29) from the fit. We then use these vectors to match resonance peaks in consecutive spectra (Supplementary Fig. 7e). For a fitted resonance peak in the transmission spectrum, we first identify all possible matching combinations in consecutive sweeps. We then use the cosine similarity metric, as defined in Supplementary Equation (30), as the cost function to find the best match. The overall resonance trace is computed by calculating the cumulative resonance shifts.

$$S_c(\hat{v}_1, \hat{v}_2) = 1 - \frac{\hat{v}_1 \cdot \hat{v}_2}{||\hat{v}_1|| \, ||\hat{v}_2||} \tag{30}$$

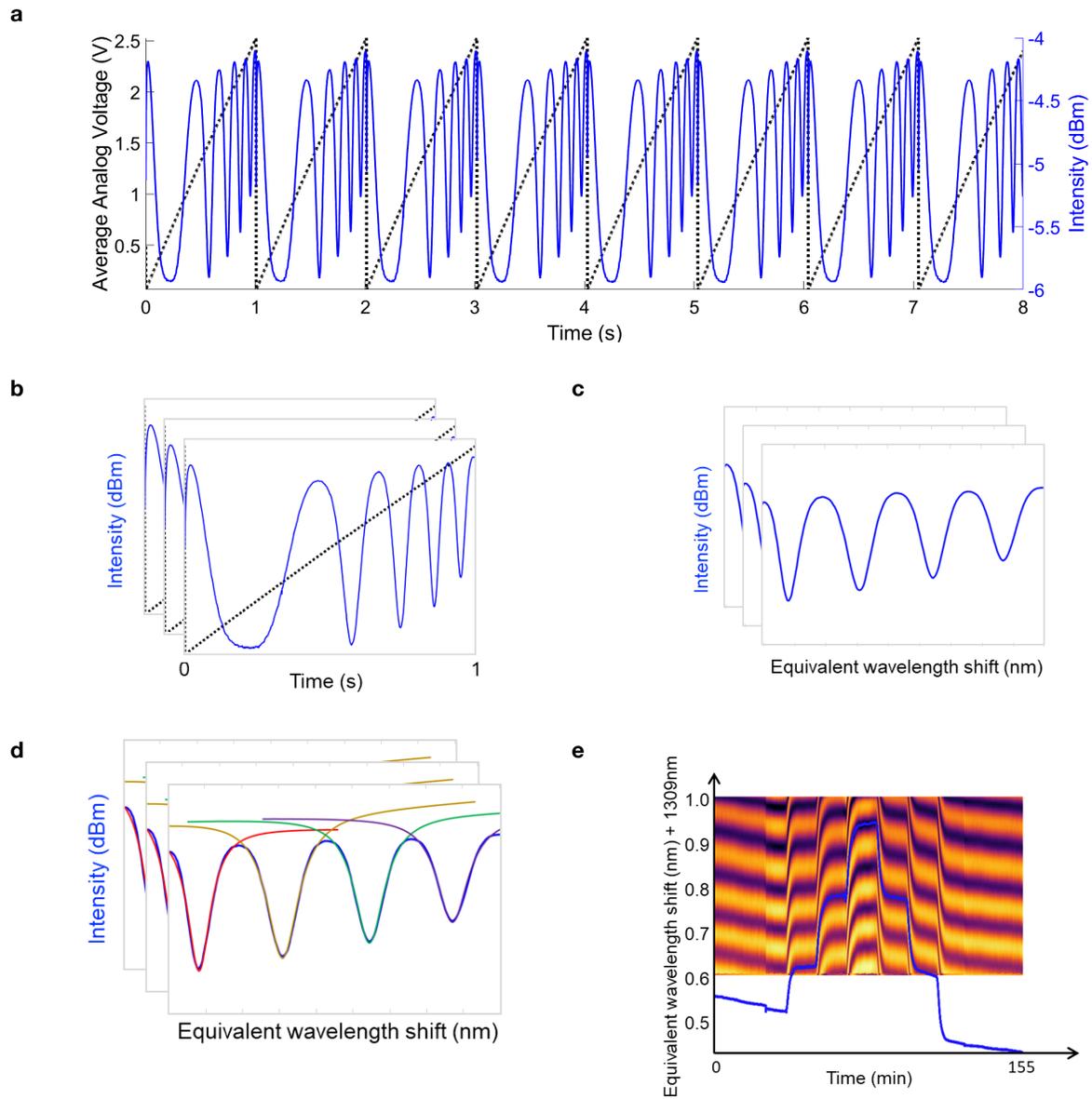

**Supplementary Figure 7. Illustration of the data analysis workflow to extract the equivalent peak shift vs. time data from fixed-wavelength sensor raw data.** (a) Optical power vs. time data are acquired, along with the tuning ramps that were used to measure the fixed-wavelength sensor resonance. (b) The data are sliced into individual pre-spectra based on the voltage ramp. (c) Individual pre-spectra are linearized into optical power vs. equivalent wavelength shift spectra as described in Supplementary Equations (26–28). (d) Resonances in individual optical spectra are identified and Lorentzian-fitted to measure the central positions and other fit parameters of each Lorentzian peak. (e) Equivalent resonance peak shifts are tracked over time (both the equivalent wavelength shift spectrogram and overlaid tracked peaks are shown). Peak shift magnitudes and dynamics introduced by bulk refractive index changes or analyte binding are quantified and plotted from these equivalent resonance peak shift vs. time data.

**Supplementary Table 1 Table of full-system cost comparison of the fixed-wavelength sensor with traditional resonator sensors using TLS readout.**

| Item Name | Item Cost (USD) | Tunable Laser Source (TLS) | | Fixed-wavelength (FW) | |
|---|---|---|---|---|---|
| | | Qty | Cost (USD) | Qty | Cost (USD) |
| Tunable laser | 39,108.88 | 1 | 39,108.88 | 0 | 0 |
| DFB laser | 1 | 0 | 0 | 1 | 1 |
| EIC | 1 | 1 | 1 | 1 | 1 |
| PIC | 1 | 1 | 1 | 1 | 1 |
| N77 module | 11,612.15 | 1 | 11,612.15 | 0 | 0 |
| InGaAs PD | 5.8 | 0 | 0 | 1 | 5.8 |
| AD2 Board | 500 | 0 | 0 | 1 | 500 |
| PCB readout electronics | 500 | 0 | 0 | 1 | 500 |
| Peltier | 33.18 | 1 | 33.18 | 1 | 33.18 |
| Aluminum plate | 20 | 1 | 20 | 1 | 20 |
| Motor controllers | 3,402.58 | 1 | 3,402.58 | 0 | 0 |
| 3-Axis motors | 2952 | 1 | 2,952 | 0 | 0 |
| Camera (top view) | 569.99 | 1 | 569.99 | 0 | 0 |
| Camera (side view) | 35 | 1 | 35 | 0 | 0 |
| Zoom lenses | 1,263.95 | 1 | 1,263.95 | 0 | 0 |
| Zoom lens extension | 870.74 | 1 | 870.74 | 0 | 0 |
| Damped optical breadboard for vibration isolation | 1,842.92 | 1 | 1,842.92 | 0 | 0 |
| Rotational micropositioner | 395 | 2 | 790 | 0 | 0 |
| Goniometer (small) | 752 | 1 | 752 | 0 | 0 |
| Goniometer (large) | 862 | 1 | 862 | 0 | 0 |
| Optical post with base | 482 | 2 | 964 | 0 | 0 |
| Post clamps | 150 | 1 | 150 | 0 | 0 |
| Cold light source | 162.33 | 1 | 162.33 | 0 | 0 |
| Total (USD) | | - | 65,394 | - | 1,062 |

$$Cost\ reduction = \frac{(Total\ cost)_{TLS} - (Total\ cost)_{FW}}{(Total\ cost)_{TLS}} \times 100\% = 98.4\ \%$$

*cost estimates obtained from publicly available new and used list pricing for components meeting the application requirements.